\documentclass[aps,floatfix,twocolumn,showpacs,showkeys]{revtex4-1}

\usepackage{epsfig}
\usepackage{graphicx}
\usepackage{amsmath,amssymb,amsopn,bm}

 \newcommand{\beq}{\begin{equation}}
 \newcommand{\eeq}{\end{equation}}
 \newcommand{\beqa}{\begin{eqnarray}}
 \newcommand{\eeqa}{\end{eqnarray}}
 \DeclareMathOperator{\im}{Im}

\begin{document}

\title{Dimuon radiation at the CERN SPS within a (3+1)d hydrodynamic+cascade 
model}
\author{E. Santini,$^{1,2}$ J. Steinheimer,$^{1,2}$ M. Bleicher$^{1,2}$, and S. Schramm$^{2}$}

\affiliation{$^{1}$Institut f\"ur Theoretische Physik, Goethe-Universit\"at, 
 Max-von-Laue-Str.~1, D-60438 Frankfurt am Main, Germany\\
 $^{2}$Frankfurt Institute for Advanced Studies (FIAS), Ruth-Moufang-Str.~1,
 D-60438 Frankfurt am Main, Germany}

\begin{abstract}
We analyze dilepton emission from hot and dense matter using a hybrid approach based on the 
Ultrarelativistic Quantum Molecular Dynamics
(UrQMD) transport model with an intermediate hydrodynamic stage
for the description of heavy-ion collisions at relativistic energies. 
During the hydrodynamic stage, the production of lepton pairs is described by
radiation rates for a strongly interacting medium in thermal equilibrium.
In the low mass region, hadronic thermal emission is 
evaluated assuming vector meson dominance including in-medium 
modifications of the $\rho$ meson spectral function through scattering from 
nucleons and pions in the heat bath. 
In the intermediate mass region, 
the hadronic rate is essentially determined by multi-pion annihilation 
processes. 
Emission from quark-antiquark annihilation in the quark gluon plasma is 
taken into account as well.
When the system is sufficiently dilute, the hydrodynamic description breaks 
down and a transition to a final cascade stage is performed. In this stage 
dimuon emission is evaluated as commonly done in transport models.
Focusing on the enhancement with respect to the contribution from long-lived 
hadron decays after freezout observed at the SPS in the low mass region of 
the dilepton spectra, the relative importance of the different thermal contributions and of the two dynamical stages is investigated. 
We find that three separated regions can be identified in 
the invariant mass spectra. Whereas the very low and the intermediate mass 
regions mostly receive contribution from the thermal dilepton emission, 
the region 
around the vector meson peak is dominated by the cascade emission. Above the 
$\rho$-peak region the spectrum is driven by QGP radiation.
Analysis of the dimuon transverse mass spectra reveals that the thermal hadronic emission shows an evident mass ordering not present in the emission from the QGP. 
A comparison of our calculation to  recent acceptance 
corrected NA60 data on invariant as well as transverse mass spectra 
is performed.

\end{abstract}

\pacs{24.10.Lx, 25.75.-q, 25.75.Dw, 25.75.Cj}
\keywords{Monte Carlo simulations, Relativistic heavy-ion collisions, Particle and resonance production, Dilepton production }

\maketitle

\section{Introduction}

Dileptons represent a penetrating probe of the 
hot and dense nuclear matter created in heavy ion collisions since, once produced, they 
essentially do not interact with the surrounding hadronic matter. 
The analysis of the electromagnetic response of the dense and hot medium 
is tightly connected to the investigation of the in-medium modification 
of the vector meson properties. Vector mesons are ideally suited for this 
exploration, because they can directly decay into a 
lepton-antilepton pair. One therefore aims to infer information on the 
modifications induced by the medium on specific properties of the vector 
meson, such as its mass and/or its width, from the invariant mass dilepton 
spectra.

The NA60 experiment at CERN SPS  has recently measured the invariant mass spectra of low mass
dimuons \cite{Arnaldi:2006jq,Arnaldi:2007ru,Arnaldi:2008fw,Arnaldi:2008gp}. As in previous ultra-relativistic heavy ion collision 
experiments 
\cite{Agakishiev:1995xb,Mazzoni:1994rb}, at low invariant masses 
an enhancement of dilepton pairs in heavy systems as compared to the expected contribution from 
decays of long-lived hadrons (mostly $\eta$, $\eta^\prime$, $\omega$ and $\phi$ mesons), 
commonly referred to as ``hadron cocktail'',
was observed. The high data quality of NA60
allowed to subtract the cocktail of the decay sources from 
the total data and to isolate the excess of pairs. 
One of the  first results was a strong evidence for the presence of broadening of the spectral 
function of the $\rho$ meson \cite{Arnaldi:2006jq}.

From the first communication by the NA60 collaboration \cite{Arnaldi:2006jq}, the remaining excess of pairs 
(often referred to as ``the excess'') 
has been object of various theoretical investigations. Phenomenological models have been refined and extended to include 
the various sources of dilepton production, and an almost 
comprehensive \cite{vanHees:2006ng,Dusling:2006yv,Renk:2006qr,Dusling:2007kh,Ruppert:2007cr,vanHees:2007th} 
interpretation of the invariant mass spectra of the 
excess has been reached. 
Calculations performed by convolution of dilepton production rates, using the assumption of local equilibrium, 
over the space-time evolution of the medium modelled according to simple
expanding fireball approaches have been quite successful in explaining the dilepton excess observed 
at SPS \cite{vanHees:2006ng,Renk:2006qr,Ruppert:2007cr,vanHees:2007th}. 

This use of basic approaches for the modelling of the heavy-ion collision 
dynamics focused the investigations on the various scenarios for the in-medium modification of the vector 
meson properties, 
and an evidence for a broadening of the $\rho$ meson spectral function was inferred from the comparison to the measurements.
However, in these models the time development of the reaction is not described by 
fully dynamical simulations but rather parametrized in terms of the estimated time which the system 
spends in the various phases. 
Though such parametrizations can be properly tuned and constrained, 
a more complete treatment of the heavy-ion dynamics in 
terms of transport and/or fully (3+1) dimensional hydrodynamical models should be in order, since a good knowledge 
of the temperature, baryon-density and flow evolution is necessary to convert dilepton rates into space-time integrated spectra.

Earlier, a (2+1) dimensional hydrodynamical description of dilepton production has been performed by 
Huoviven et al. \cite{Huovinen:1998tq, Huovinen:1998ze,Huovinen:2002im} and compared to data from the first generation 
of dilepton experiments performed 
in the nineties. However, already at that time \cite{Huovinen:2002im} it was shown that the evolution can have strong impact on the dilepton yields. To our knowledge, the model was never applied to the high resolution NA60 data.
Application of hydrodynamics to analyse the dimuon excess observed by NA60 were performed by 
Dusling et al.  \cite{Dusling:2006yv,Dusling:2007kh} using a boost invariant 
(1+1) hydrodynamical model. If on  one side this represents a step further towards the 
inclusion of a more realistic dynamics, on the other side boost invariant hydrodynamics is expected to 
be an approximation more valid at RHIC energies than at SPS. 
In Ref. \cite{Huovinen:2001wx}, e.g., it was shown that the photon spectra measured by the WA98 
Collaboration at the CERN SPS  \cite{Aggarwal:2000th} could be accounted for in a boost-invariant 
scenario only by choosing a short initial time ($\tau_0$=0.3 fm) which is not in accordance with the 
expected longitudinal geometry. 
In this respect, it is surely desirable to relax this approximation and use a full
(3+1) hydrodynamic expansion. 

Moreover, the contribution to the excess of non-thermal dilepton radiation from short 
living mesons has only very recently 
received proper attention by the theoretical groups mostly active in the field of thermal dileptons. 
For example, in Ref.  \cite{Dusling:2006yv} the contribution from freezeout $\rho$ mesons was 
neglected and added only in a later work  \cite{Dusling:2007kh}, whereas in Ref.~\cite{vanHees:2007th} 
an explicit treatment of $\rho$ decays at 
thermal freezeout together with the additional inclusion of a non-thermal 
component of primordially produced $\rho$ 
mesons which escape the fireball were required in order to account for 
some discrepancy with thermal emission revealed by analysis of measured 
transverse pair momenta ($p_T$) spectra at $p_T$$>$1 GeV. 

In this paper, we want to directly address these aspects presenting a 
consistent calculation 
of the dilepton production at SPS energy within a model which attempts to take 
into account  both the complexity of the dilepton rate in hot a dense medium 
as well as the complexity of the pre-, post-, and equilibrium 
heavy-ion dynamics. 
The latter is modelled with an integrated 
Boltzmann+hydrodynamics hybrid approach based on the Ultrarelativistic 
Quantum Molecular Dynamics (UrQMD) transport model with an intermediate 
(3+1) dimensional ideal hydrodynamic stage. For earlier investigations within hybrid approaches see e.g. \cite{Dumitru:1999sf,Bass:1999tu,Hirano:2005xf}. The hybrid approach used here has 
been successfully applied to many bulk observables and photon spectra 
\cite{Bauchle:2009ep} at SPS energies.
For the present investigation, during the local equilibrium phase, the 
radiation rate of the strongly 
interacting medium is standardly 
modelled using the vector meson dominance model and related to the spectral 
properties of the light vector mesons, with the $\rho$ meson having the 
dominant role.
In-medium modifications of the $\rho$-meson spectral function due to 
scattering from hadrons in the heat bath are properly included in the model.
Two additional sources of thermal radiation, namely  emission 
from multi-pion annihilation processes and from a thermalized partonic 
phase are included as well.

The paper is organized as follows. In Sec. \ref{The model}, we briefly 
discuss the hybrid model and present the emission rates of the 
various sources taken into account.
In Sec. \ref{Results} we present results for the dilepton excess 
invariant mass spectra and transverse mass spectra.  A comparison to 
the NA60 data is performed. 
Section \ref{decomp} is dedicated to a discussion of the the 
various components 
that enter the contribution of the cascade stage of the hybrid approach.
Finally, a summary and conclusions are given in Sec. \ref{Conclusions}.

\section{The model \label{The model}}

\subsection{The hybrid approach}

To simulate the dynamics
of the In+In collisions we employ a transport approach with an
embedded three-dimensional ideal relativistic one fluid
evolution for the hot and dense stage of the reaction
based on the UrQMD
 model. The present hybrid approach has 
been extensively described 
in  Ref. \cite{Petersen:2008dd} and first applications to e.m. probes have 
been recently 
performed \cite{Bauchle:2009ep,Santini:2009nd,Bauchle:2010ym,Bauchle:2010yq,Santini:2010in}. Here, we limit ourselves 
to briefly describe  its main features and refer the reader to Ref. 
\cite{Petersen:2008dd} for details.

UrQMD \cite{Bass:1998ca,Bleicher:1999xi,Petersen:2008kb} is a hadronic 
transport 
approach which simulates multiple 
interactions of ingoing
and newly produced particles, the excitation and fragmentation
of color strings and the formation and decay
of hadronic resonances.  
The coupling between the
UrQMD initial state and the hydrodynamical evolution
proceeds when the two Lorentz-contracted nuclei have
passed through each other. Here, the spectators continue to propagate 
in the cascade and all other hadrons
are mapped to the hydrodynamic grid. Event-by-event fluctuations are 
directly taken into account via initial conditions
generated by the primary collisions and string fragmentations in the 
microscopic UrQMD model. This leads to
non-trivial velocity and energy density distributions for
the hydrodynamical initial conditions in each single event~\cite{Steinheimer:2007iy,Petersen:2009vx}. 
Subsequently, a full (3+1) dimensional
ideal hydrodynamic evolution is performed using
the SHASTA algorithm \cite{Rischke:1995ir,Rischke:1995mt}. The hydrodynamic
evolution is gradually merged into the hadronic cascade: 
to mimic an iso-eigentime hypersurface, full
transverse slices, of thickness $\Delta z$ = 0.2fm, are transformed to particles whenever in all
cells of each individual slice the energy density  drops below
five times the ground state energy density. 
The employment of such gradual transition allows to 
obtain a rapidity independent transition temperature 
without artificial time dilatation effects 
\cite{Petersen:2009gu,Steinheimer:2009nn} and has 
been explored in detail in various recent works 
\cite{Li:2008qm,Petersen:2009mz,Petersen:1900zz,Petersen:2009gu} 
devoted to SPS conditions. 
When merging, the hydrodynamic fields are transformed to particle
degrees of freedom via the Cooper-Frye equation \cite{Cooper:1974mv} on the hypersurface $\sigma_\mu$
\begin{equation}
\label{cooper_frye}
p_0 \frac{dN}{d^3p}=\int_\sigma f(x,p) p^\mu d\sigma_\mu\,,
\end{equation}
where $f(x,p)$ are the boosted Fermi or Bose distributions 
corresponding to the respective particle species.
For the present analysis,
the latter was modified in order to account for the 
spectral  shape of the $\rho$ meson via the substitution:
\beq
\frac{d^3p}{p_0}\rightarrow \frac{d^3p}{p_0}dM^2\delta^+(M^2-m_\rho^2)\rightarrow d^4p 2A_\rho(M)\,,
\eeq
with $A_\rho(M)=-\frac{1}{\pi}\im D_\rho(M)$ being the $\rho$ meson spectral 
function and $D_\rho(M)$ being the $\rho$ meson propagator in vacuum. 
The created particles then proceed in their evolution in
the hadronic cascade where rescatterings and
final decays occur until all interactions cease and
the system decouples.

An input for the hydrodynamical calculation is
the equation of state (EoS). In this work we employ 
an equation of state in line with lattice data that 
follows from coupling the Polyakov loop to a chiral 
hadronic flavor-SU(3) model \cite{Steinheimer:2010ib}. 
The hadronic part is  an extension of a non-linear representation of a
sigma-omega model including the lowest-lying multiplets
of baryons and mesons (for the derivation and a detailed
discussion of the hadronic part of the model Lagrangian see 
\cite{Papazoglou:1997uw,Papazoglou:1998vr,Dexheimer:2008ax}). 
In spirit similar to the PNJL model \cite{Fukushima:2003fw} it includes the 
Polyakov loop as an effective field and adds quark degrees of freedom.
In this configuration the EoS describes chiral restoration as well as the 
deconfinement phase transition, while it contains the correct asymptotic 
degrees of freedom (quarks $\leftrightarrow$ hadrons). For details, we refer the reader to 
Ref.~\cite{Steinheimer:2010ib}

\subsection{ Thermal contributions to the dimuon excess: emission rates }
During the locally equilibrated hydrodynamical stage, dimuon  emission is 
calculated locally in space-time, i.e. for each cell of the (3+1) hydrodynamical 
grid, according to the expression for
 the thermal equilibrium rate of dilepton emission 
per four-volume and
four-momentum from a bath at temperature $T$ and baryon chemical
potential $\mu_B$. 
The total dimuon yield  is  then
obtained integrating  the emission rate over all fluid cells and all time 
steps of the (3+1) grid that are spanned by the system during the 
hydrodynamical evolution until the transition criterium is reached. 

In the low invariant mass region of the dilepton 
spectrum ($M<1$ GeV) the 
largest contribution to the dilepton excess is due to the 
$\rho^0\rightarrow l^+l^-$ emission.
As pointed out in various works 
(see e.g. \cite{Gale:1990pn,Rapp:1999ej,Ruppert:2007cr,vanHees:2007th}), 
when invoking the  vector meson dominance model, the thermal rate can be related to the 
spectral properties of the vector mesons. 
Retaining only the contribution of the $\rho$ meson (which is the largest), 
one arrives at the following expression  \cite{Rapp:1999ej}:
\beq
\frac{d^8 N_{\rho\rightarrow ll}}{d^4 x d^4 q}=-\frac{\alpha^2m_\rho^4}{\pi^3 g_\rho^2}\frac{L(M^2)}{M^2}f_B(q_0;T)
\im D_\rho(M,q;T,\mu_B) \, ,  
\label{rate}
\eeq
where $\alpha$=$e^2/(4 \pi)$=1/137 denotes the fine structure constant, 
$M^2=q_0^2-q^2$ the dilepton invariant mass squared with energy $q_0$
and three-momentum $q$, $f_B(q_0;T)$ the thermal Bose
distribution function, $L(M^2)$ the lepton phase space factor,
\begin{equation}
\label{lept-ps}
L(M^2)=\left (1+\frac{2 m_l^2}{M^2} \right) \sqrt{1-\frac{4 m_l^2}{M^2}} 
\ , 
\end{equation}
that quickly approaches one above threshold, and $\im D_\rho(M,q;T,\mu_B)$ 
the imaginary part of the in-medium $\rho$ meson propagator
\beq
D_\rho(M,q;T,\mu_B)=\frac{1}{M^2-m_\rho^2-\Sigma_\rho(M,q;T,\mu_B)}\, .
\label{rhoprop}
\eeq
Here, $m_\rho$ denotes the pole mass and $\Sigma_\rho(M,q;T,\mu_B)$ the 
in medium self-energy. In this application, the self-energy
contributions taken into account are the following: 
\beq
\Sigma_\rho(M,q;T,\mu_B)=\Sigma^0(M)+\Sigma^{\rho\pi}(q;T)+
\Sigma^{\rho N}(q;T,\mu_B)\, , 
\label{selfen}
\eeq
where $\Sigma^0(M)$ is the vacuum self-energy and $\Sigma^{\rho\pi}(q;T)$ and $\Sigma^{\rho N}(q;T,\mu_B)$
denote the contribution to the self-energy due to the direct
interactions of the $\rho$ with, respectively, pions and nucleons of the
surrounding heat bath.
The latter have been calculated according to Ref.~\cite{Eletsky:2001bb}, 
where they were evaluated in terms of empirical scattering amplitudes from 
resonance dominance at low energies and Regge-type behaviour at high energy. 
In principle, the matter part of the self-energy should  depend on 
$M$ and $q$ (or on $q^0$ and $q$) separately. Here, it depends only on $q$ 
because in Ref.~\cite{Eletsky:2001bb} the scattering amplitudes were 
evaluated on the mass shell of the $\rho$ meson.

Our analysis of the invariant mass dilepton spectrum is restricted to $M$$<$1.5 GeV. 
In the mass region 1$<$$M$$<$1.5 GeV
the sources expected to give a major contribution to the dilepton excess 
are the multi-pion emission and the $q \bar q$ annihilation in the QGP. 
A third background source usually present in the intermediate mass region, 
the correlated decays of $D$ and $\bar{D}$ mesons, 
could be disentangled  from the prompt excess via vertex reconstruction 
\cite{Arnaldi:2008er} and have been recently subtracted from the NA60 data 
on the excess \cite{Arnaldi:2008fw}.

The four-pion contribution is the  major manifestation of the
 multi-pion emission for the region of our interest, 1$<$$M$$<$1.5 GeV, and 
 will be included in this analysis.  
Six-pion contribution as well as other multi-hadron interactions are not included 
since they start to play a non-negligible role only for masses $M$$\gtrsim$1.8 GeV.
Following Huang's observation \cite{Huang:1995dd} that lepton pair production can be determined from the electromagnetic spectral function extracted in $e^+e^-$ annihilation, 
we estimate the contribution  of the $4\pi\rightarrow l^+l^-$ process to 
the dilepton rate as
\beq
\frac{d^8 N_{4\pi\rightarrow ll}}{d^4 x d^4q}= \frac{4\alpha^2}{(2\pi)^2}e^{-q_0/T}\frac{M^2}{16\pi^3\alpha^2}\sigma(e^+e^-\rightarrow 4\pi)\,.
\label{rate_4pi}
\eeq
The above expression neglects the soft-pion final state interaction corrections that induce the mixing between the axial-vector and vector currents. We postpone the investigation of the effects caused by the parity mixing phenomenon to a future work and, here, restrict ourselves to mention that the mixing is expected to enhance the rate in the invariant mass region 1$<$$M$$<$1.3 GeV \cite{Huang:1995dd}. Calculations that include the axial vector-vector mixing can be found in Refs. \cite{vanHees:2006ng,vanHees:2007th}.

For the specific purposes of this work, it is sufficient to use an empirical approach and estimate the four-pion contribution to the dilepton spectra directly from the measured cross sections, without attempting any phenomenological modelling of the underlying processes.  
We consider the multi-pion reactions 
$e^+e^-\rightarrow \pi^+\pi^-\pi^+\pi^-$ and $e^+e^-\rightarrow\pi^+\pi^-\pi^0\pi^0$. 
 As a side remark, we  mention that models based on effective Lagrangians suggest that these reactions are dominated by processes involving the two-body  intermediate state $a_1(1260)\pi$ \cite{Song:1994zs,Lichard:2006kw}. 
For the cross section of the $e^+e^-\rightarrow \pi^+\pi^-\pi^+\pi^-$ process we use the 
recent precise  BaBar data \cite{Aubert:2005eg} which cover a  large range of center of mass energies . The measured cross section is in good agreement with
the high precision data taken at VEPP-2M by SND \cite{Achasov:2003bv} and
CMD-2 \cite{Akhmetshin:1999ty,Akhmetshin:2004dy} in the energy range 0.7--1.4 GeV, as well
as with data obtained at DCI by DM2 \cite{Bisello:1990du} in 1.4--2.0 GeV
range. 
In a similar spirit, for the cross section of the process $e^+e^-\rightarrow\pi^+\pi^-\pi^0\pi^0$, 
we  take preliminary results of the BaBar Collaboration \cite{Druzhinin:2007cs}, 
which agree with SND \cite{Achasov:2003bv} measurements in  the energy range below 1.4 GeV.

Finally, the $q \bar q \to \gamma^* \to l^+l^-$ emission is evaluated according to ~\cite{Cleymans:1986na} as
\begin{widetext}
\begin{equation} 
\frac{dN_{q\bar q \to ll}}{d^4 x d^4q} = \frac{\alpha^2}{4\pi^4} 
\frac{T}{q} f_B(q_0;T) \sum\limits_q e_q^2 \ \ln
\frac{\left(x_-+\exp[-(q_0+\mu_q)/T]\right) \left( x_++\exp[-\mu_q/T]\right)}
{\left(x_++\exp[-(q_0+\mu_q)/T]\right) \left( x_-+\exp[-\mu_q/T]\right)} \,,
\label{qqrate2} 
\end{equation} 
\end{widetext}
with $x_\pm=\exp[-(q_0\pm q)/2T]$ and $\mu_q$ the quark chemical potential.    

Eqs.~(\ref{rate}), (\ref{rate_4pi}) and (\ref{qqrate2}) are valid in the rest frame of the fluid. For a moving 
fluid the Bose/Boltzmann distribution function must be substituted with the 
J\"{u}ttner function. The modulus of the three momentum can be expressed as a function 
of the Lorentz invariant variables $M$ and $q_\mu u^\mu$ 
($u^\mu$ denotes the four-velocity of the fluid cell) as $\sqrt{(q_\mu u^\mu)^2-M^2}$, 
so that, e.g., the in-medium $\rho$ meson propagator in Eq.~(\ref{rate}) can be expressed 
as $D_\rho(M,\sqrt{(q_\mu u^\mu)^2-M^2};T,\mu_B)$. In brief, the substitution 
$(q_0,q)\rightarrow (q_\mu u^\mu,\sqrt{(q_\mu u^\mu)^2-M^2})$ must be performed on the r.h.s. of Eqs.~(\ref{rate}), (\ref{rate_4pi}) and (\ref{qqrate2}).

\begin{figure}
\includegraphics[width=.45\textwidth]{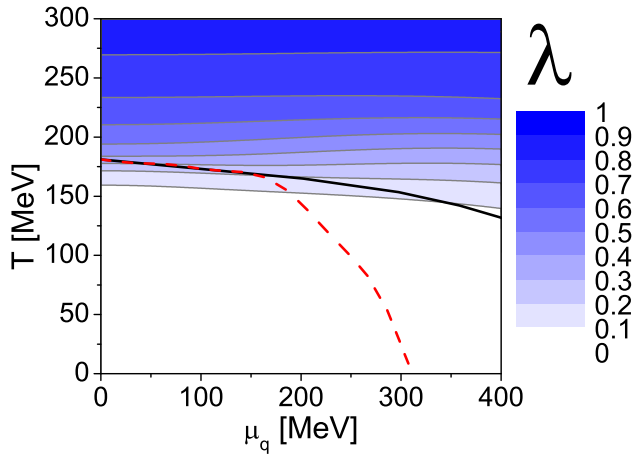}
\caption{(Color online) Fraction of QGP for various values of temperature and quark chemical potential. The dashed line indicates where the change of
the chiral condensate with respect to $T$ and $\mu_q$ has a maximum 
while the solid line shows the same for the change of the Polyakov loop. See Ref.~\cite{Steinheimer:2010ib} for details.
 \label{lambda}}
\end{figure}

The $\chi$-EoS used in this analysis includes an extended crossover between 
the hadronic and the QGP phase during which the two states of matter coexist. 
The fraction of QGP in the medium is estimated as 
the ratio between the energy density stored in quarks and gluons degrees of 
freedom and the total energy density. We denote this ratio as 
$\lambda=\lambda(T,\mu_q)$. Typical values of $\lambda$ as a function of temperature 
$T$ and quark 
chemical potential $\mu_q$ are shown in Fig.~\ref{lambda}.
For a given configuration of $(T,\mu_q)$ of a fluid cell, we weight 
the hadronic radiation rate with the function $(1-\lambda)$ and the 
$q\bar{q}$ annihilation rate with the function $\lambda$. 
Thus, each cell contributes to the total dimuon emission rate according to:
\begin{widetext}
\begin{equation} 
\frac{d^8 N_{ll}}{d^4 x d^4q} =\left[1-\lambda(T,\mu_q)\right]\left(\frac{d^8 N_{4\pi\rightarrow ll}}{d^4 x d^4q}+\frac{d^8 N_{\rho\rightarrow ll}}{d^4 x d^4q}\right)
+ \lambda(T,\mu_q) \frac{dN_{q\bar q \to ll}}{d^4 x d^4q}\,.
\end{equation} 
\end{widetext}

\subsection{Cascade contribution to the dimuon excess }

In  the evolution stages that precede or follow the 
hydrodynamic phase, dimuon emission from the 
$\rho$-meson is calculated as in Ref.~\cite{Schmidt:2008hm,Santini:2009nd} 
employing the time integration method  (often called also ``shining method'') 
that has since long been 
applied in the transport description of dilepton emission 
(see e.g. \cite{Li:1994cj,Vogel:2007yu,Bratkovskaya:2007jk,Schmidt:2008hm}).
Note that in the pre- and post-hydrodynamical stages the particles are the 
explicit degrees of freedom and all their interactions are treated 
explicitly
within the cascade transport approach.  
This allows to dynamically account for final state interactions during the 
late stage of the reaction and to follow 
the continuous decoupling of the different particles, since 
sequential freezeout of different particle species occurs 
depending on the microscopic reaction rates. 
For such a minutely detailed microscopic description of the evolution of 
the system, however, there is a price to pay, namely 
that the cascade model
solely treats collisions and decays on the basis
of vacuum cross sections and decay rates.
Thus, eventual residual in-medium modifications of the 
$\rho$ meson spectral function in this last stage will be 
neglected when adopting the present model. 
In fact, the consistent treatment of broad spectral structures in transport 
approach
is not trivial. Many works have been dedicated to this topic, e.g. 
Refs.~\cite{Knoll:1998iu,Leupold:1999ga,Ivanov:1999tj,Cassing:2000ch,Knoll:2001jx,Juchem:2003bi,Juchem:2004cs}; 
for an overview we refer the reader to the recent 
critical review by J.~Knoll~\cite{CBM:2010}. 

Emission from the stage preceding the hydrodynamical evolution is typically small, since 
the geometrical criterium adopted to start the hydrodynamical evolution corresponds to 
a starting time $t_{\rm start}\approx 1.16$ fm at top SPS energy. Emission from the stage that follows the hydrodynamical evolution receives two main contributions: when merging the hydrodynamical stage to the UrQMD model to perform the final cascade, 
the hydrodynamic fields are mapped to hadrons according to the Cooper-Frye equation. At this point a certain number of primary $\rho^0$'s are created and enter the cascade. If soon after the transition the system is decoupled with respect to processes involving $\rho$ mesons, during the cascade these primary $\rho^0$ mesons simply decay, 
no further $\rho^0$'s are generated and the corresponding dilepton yield is 
determined by the abundance of $\rho^0$ created  at the transition 
times the dilepton branching.
If the system is not decoupled  with respect to processes involving $\rho$ mesons, as it is presumably in reality and in the present model (as we will show), $\rho$ meson (re)generation and absorption will occur through processes such as $\pi\pi$ annihilation and 
resonance decays.  These processes will delay the decoupling, increase the emission time and, consequently, the dilepton yield.

\section{Comparison to NA60 data \label{Results}}
\subsection{Centrality selection}
\begin{figure}
\includegraphics[width=.45\textwidth]{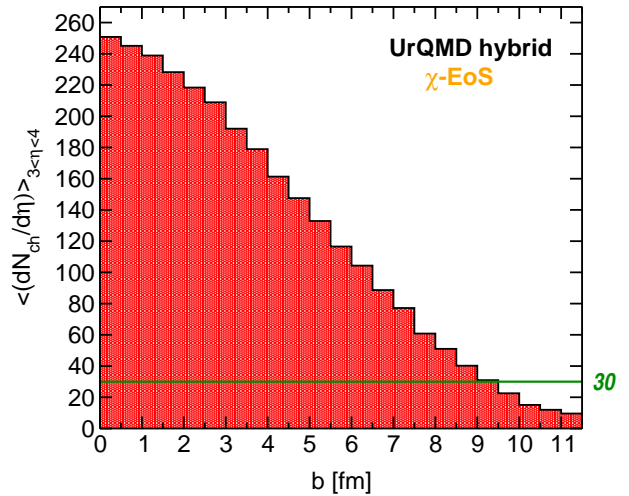}
\caption{(Color online) Charged particle density as a function of the impact parameter. \label{ch_part}}
\end{figure}

The NA60 Collaboration has recently presented data fully corrected for 
geometrical acceptance and pair efficiencies of the NA60 detector 
\cite{Arnaldi:2008fw}.
The acceptance-corrected data correspond to nearly minimum bias collisions, 
selecting events with a charged particle density $dN_{ch}/d\eta$$>$30.
In order to select the appropriate impact parameter range in our 
simulations, we first simulate minimum bias collisions and determine the 
charged particle density as a function of the impact parameter. 
The result is shown in Fig. \ref{ch_part}. We find that $dN_{ch}/d\eta$$>$30 
corresponds to $b$$<$9 fm.
With this selection, we obtain an average  charged particle density 
$\langle dN_{ch}/d\eta \rangle$$=$115, value that deviates from the 
measured one $\langle dN_{ch}/d\eta \rangle$$=$120 only by 4\%.
\subsection{Invariant mass spectra}
\begin{figure*}[t]
\includegraphics[width=.295\textwidth,angle=-90]{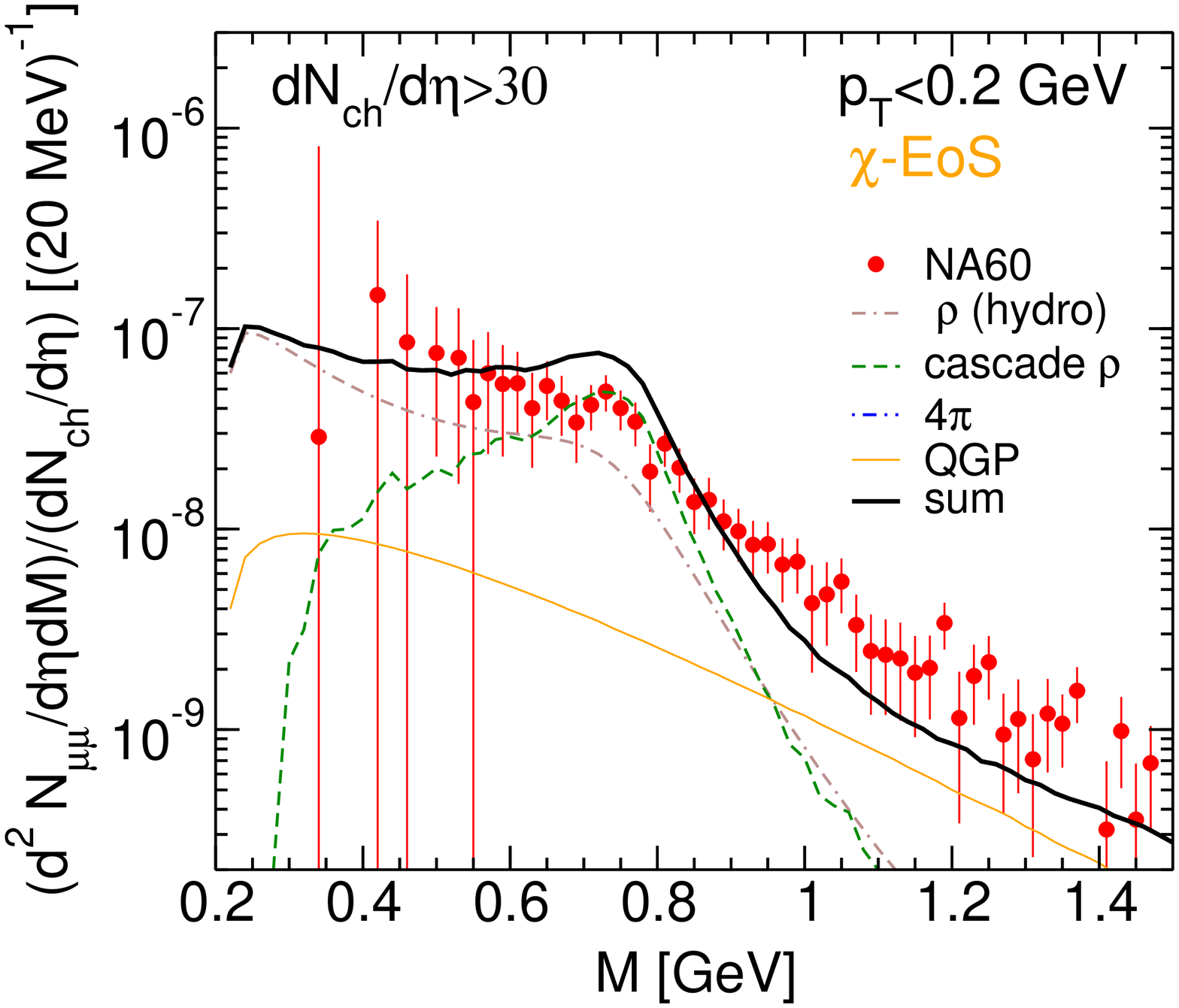}\hspace{-1.4cm}
\includegraphics[width=.295\textwidth,angle=-90]{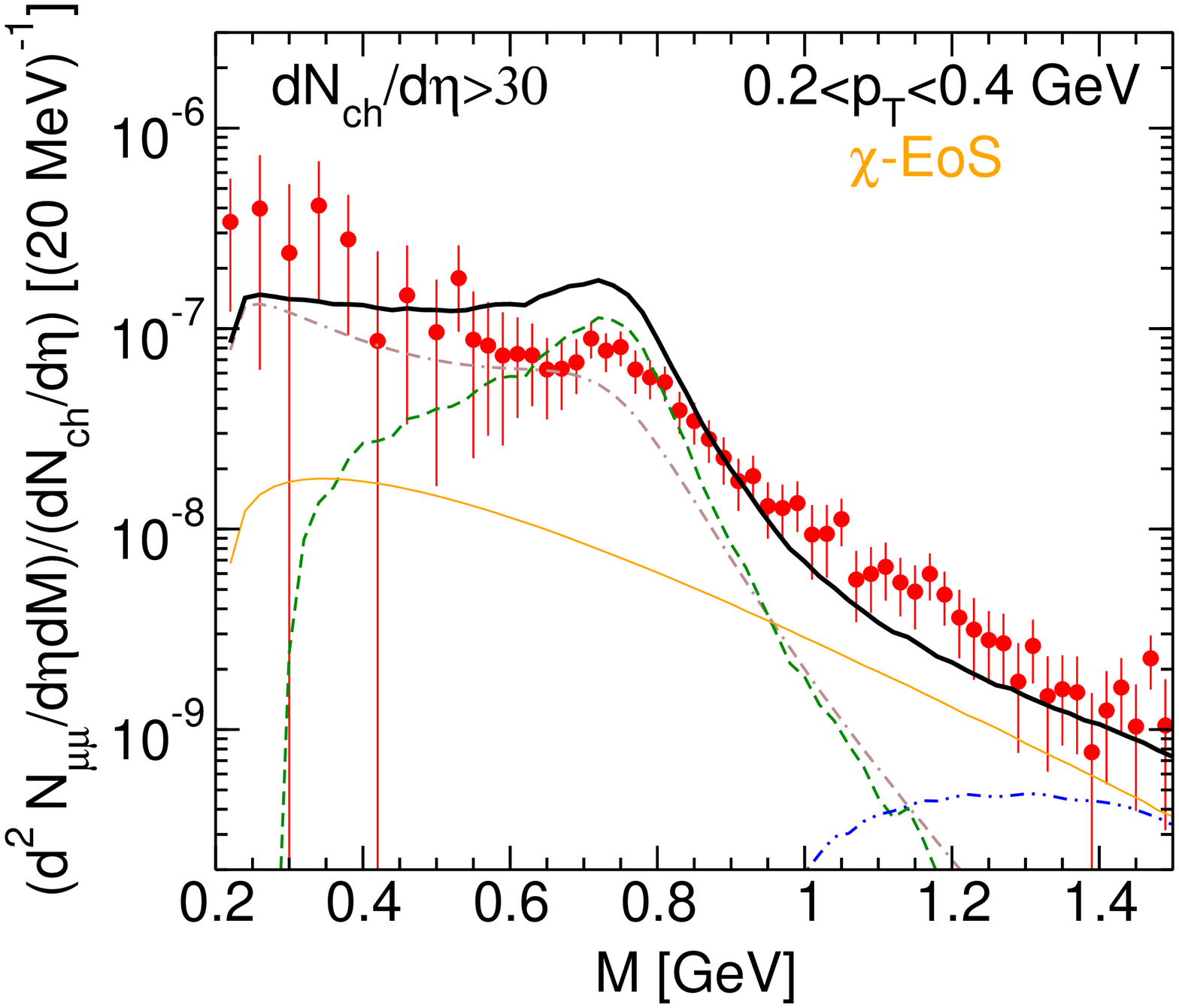}\hspace{-1.4cm}
\includegraphics[width=.295\textwidth,angle=-90]{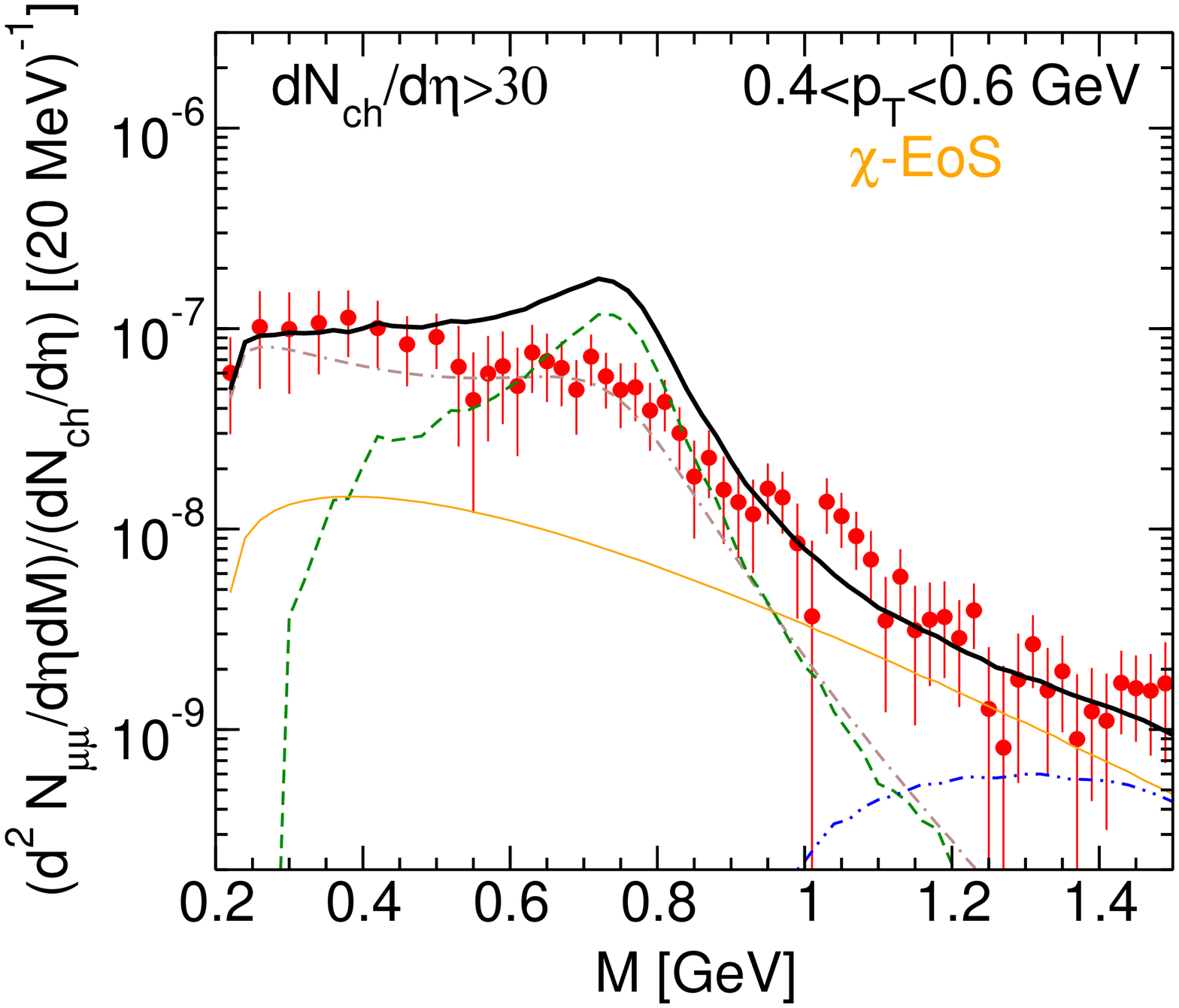}
\\
\noindent
\includegraphics[width=.295\textwidth,angle=-90]{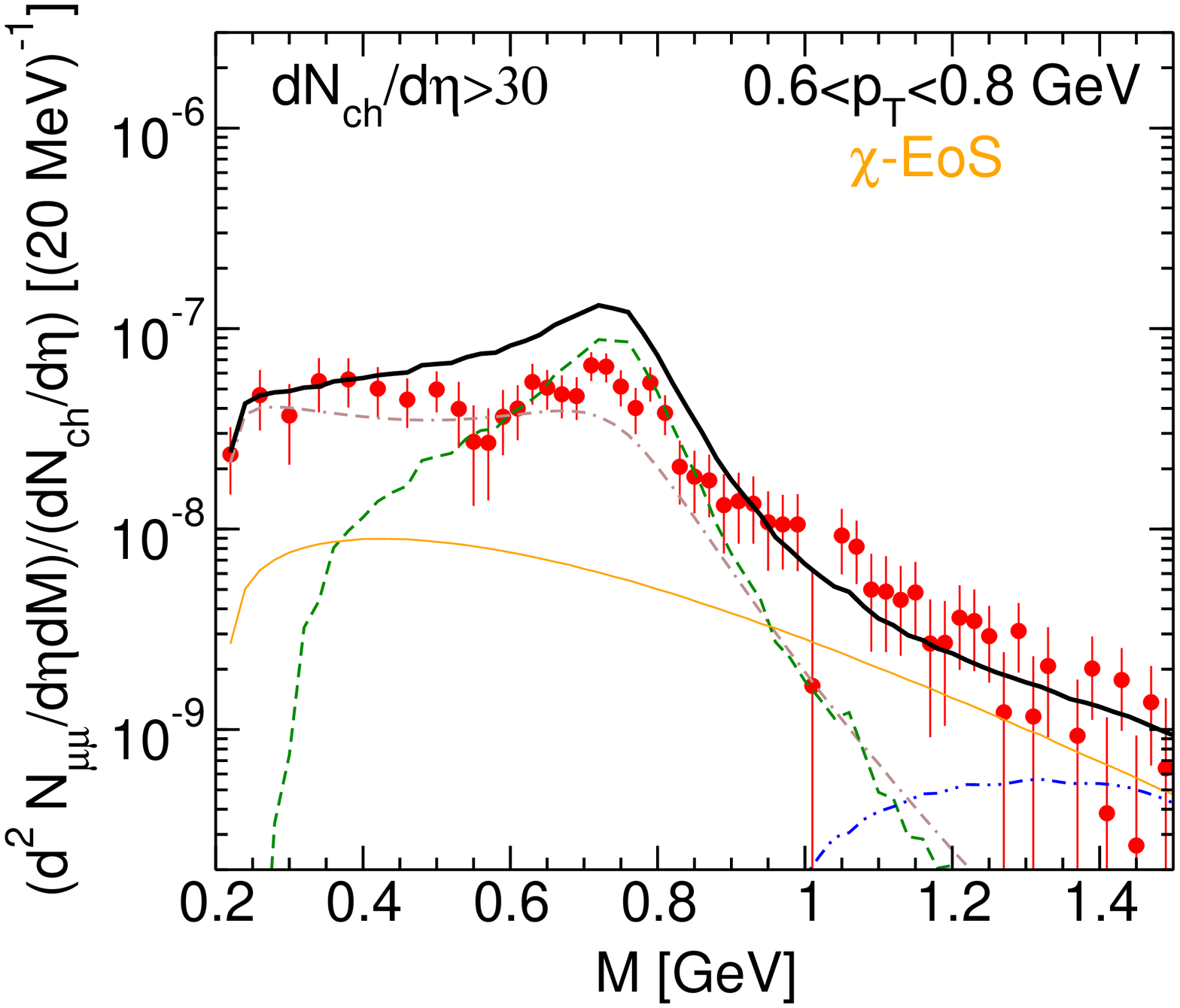}\hspace{-1.4cm}
\includegraphics[width=.295\textwidth,angle=-90]{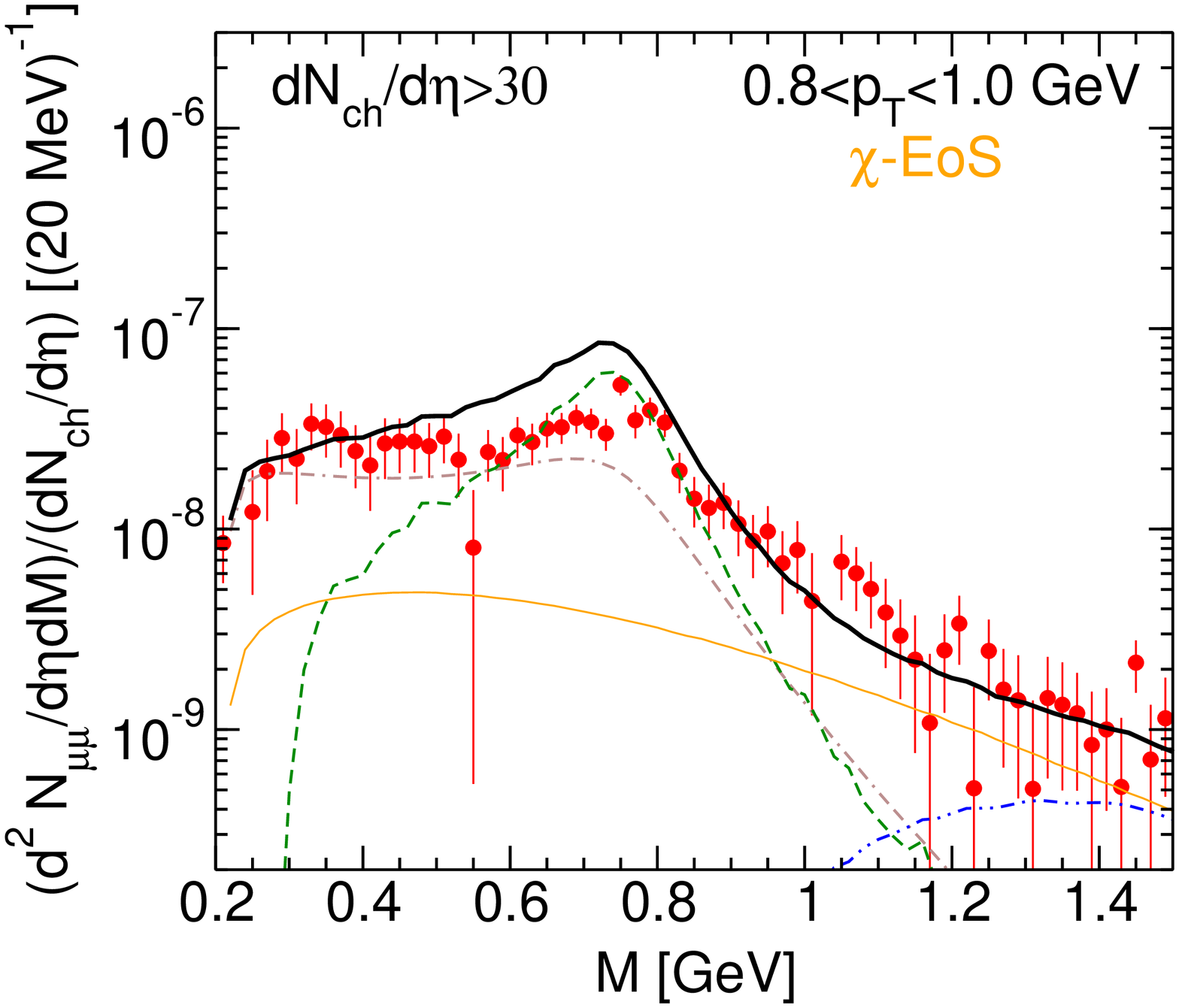}\hspace{-1.4cm}
\includegraphics[width=.295\textwidth,angle=-90]{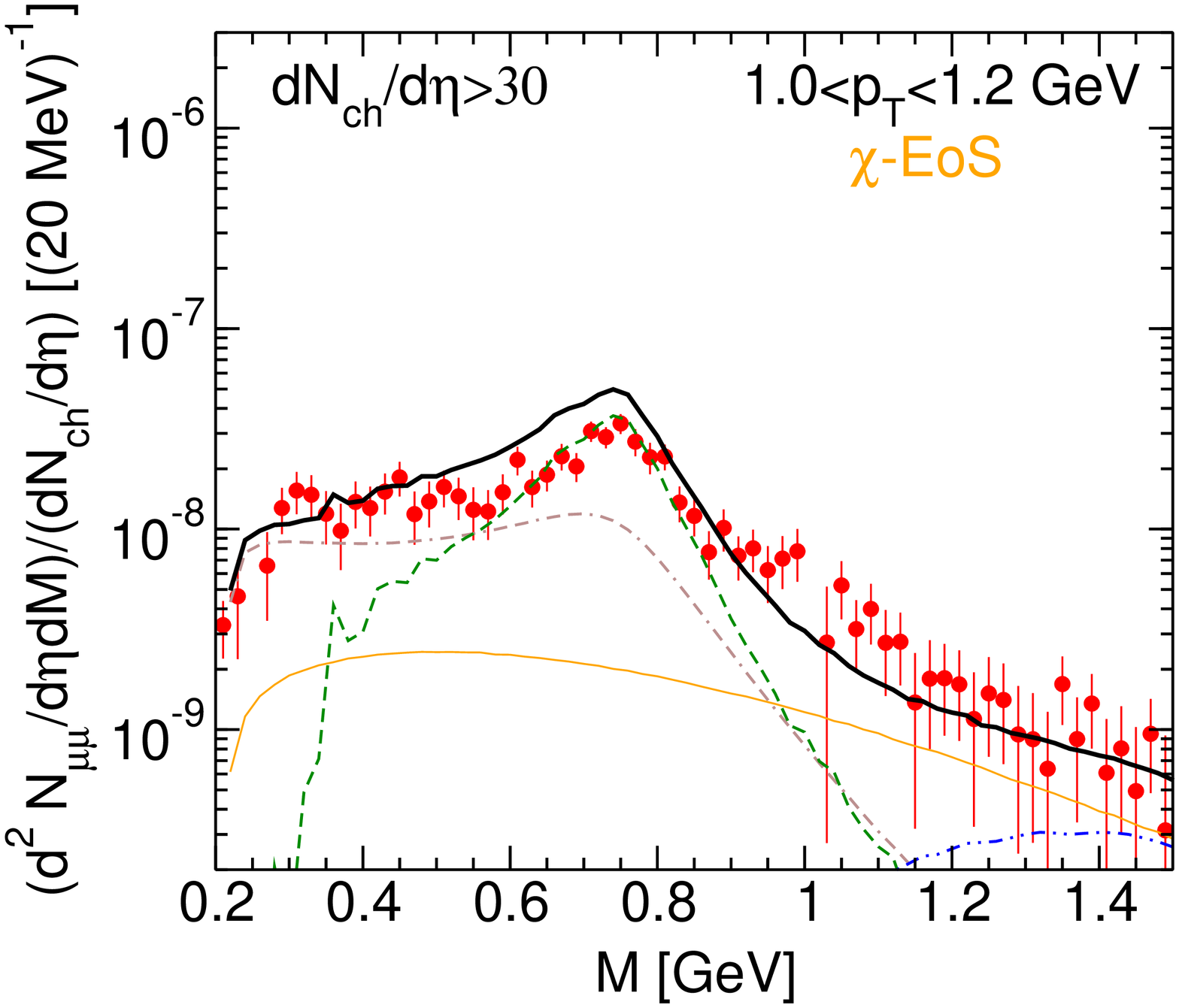}
\\
\noindent
\includegraphics[width=.295\textwidth,angle=-90]{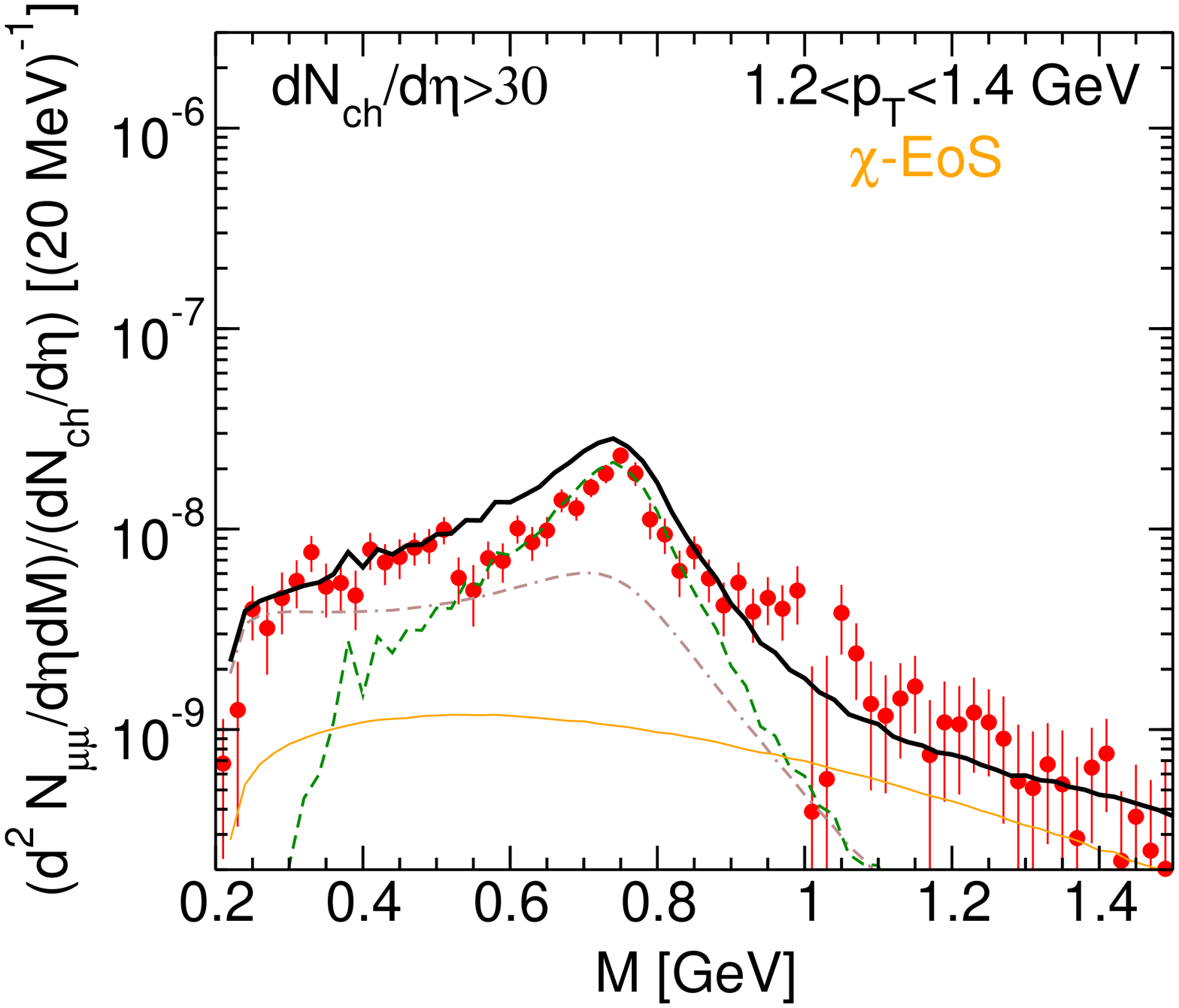}\hspace{-1.4cm}
\includegraphics[width=.295\textwidth,angle=-90]{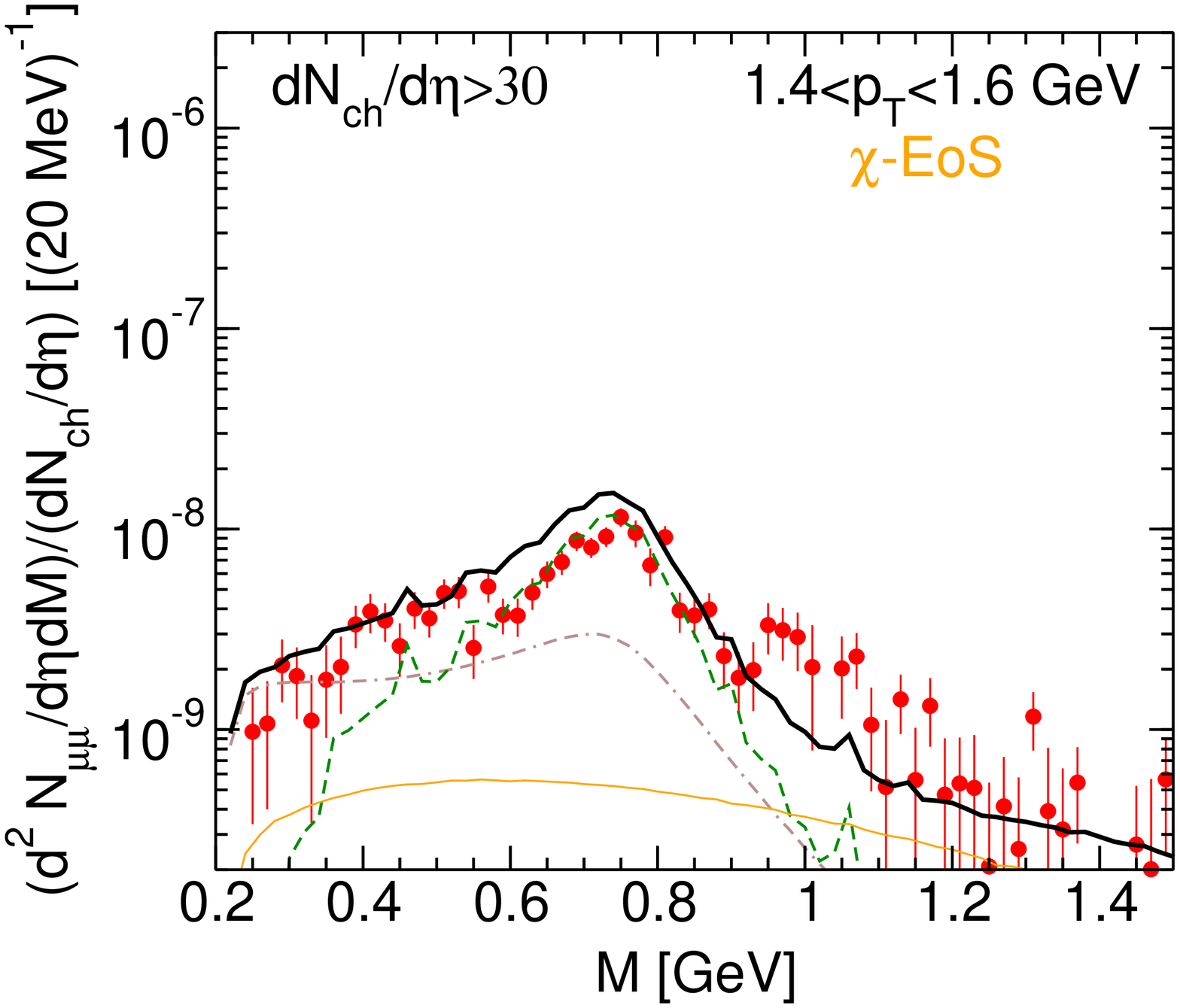}\hspace{-1.4cm}
\includegraphics[width=.295\textwidth,angle=-90]{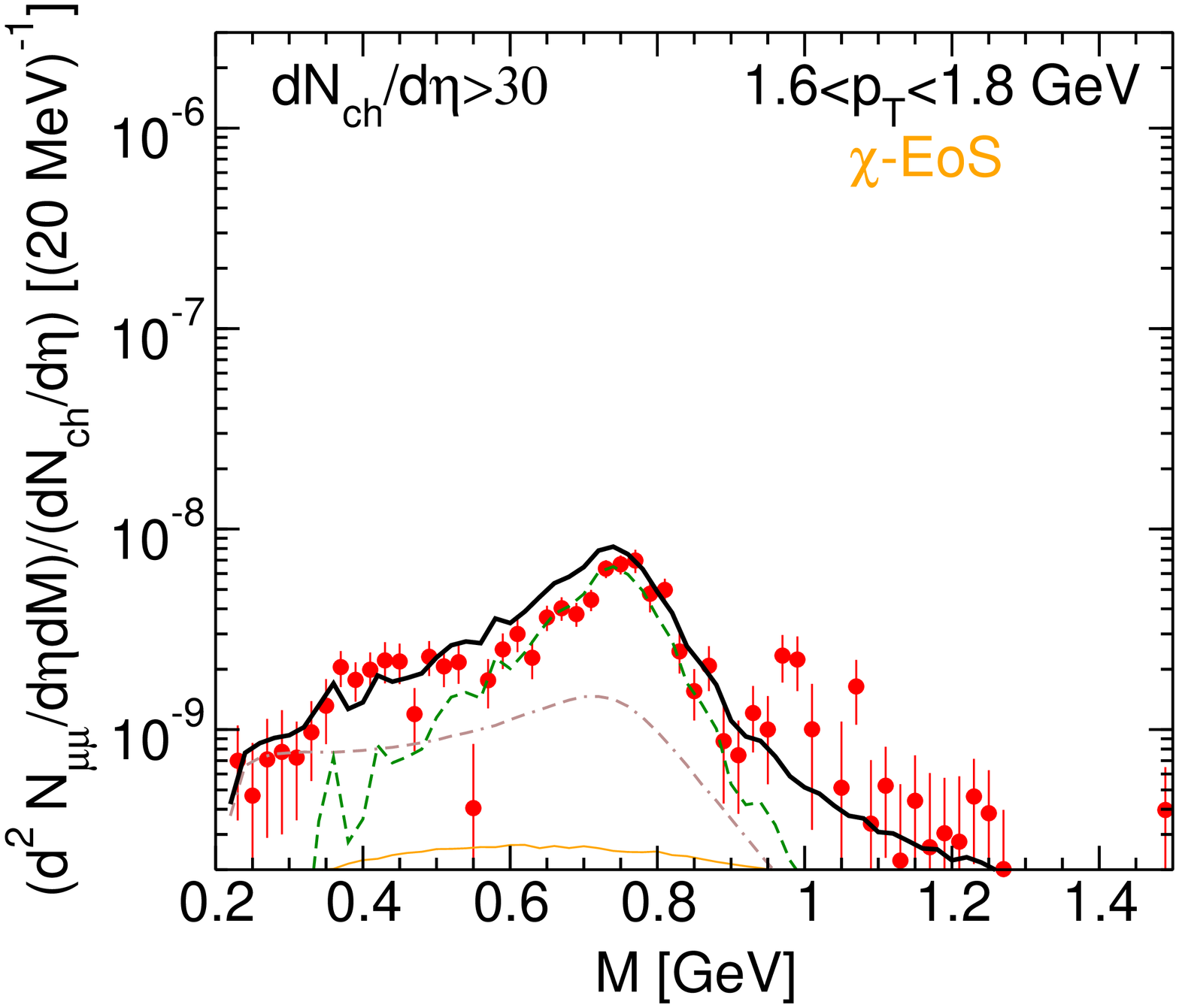}
\\
\noindent
\hspace{-0.3cm}
\includegraphics[width=.295\textwidth,angle=-90]{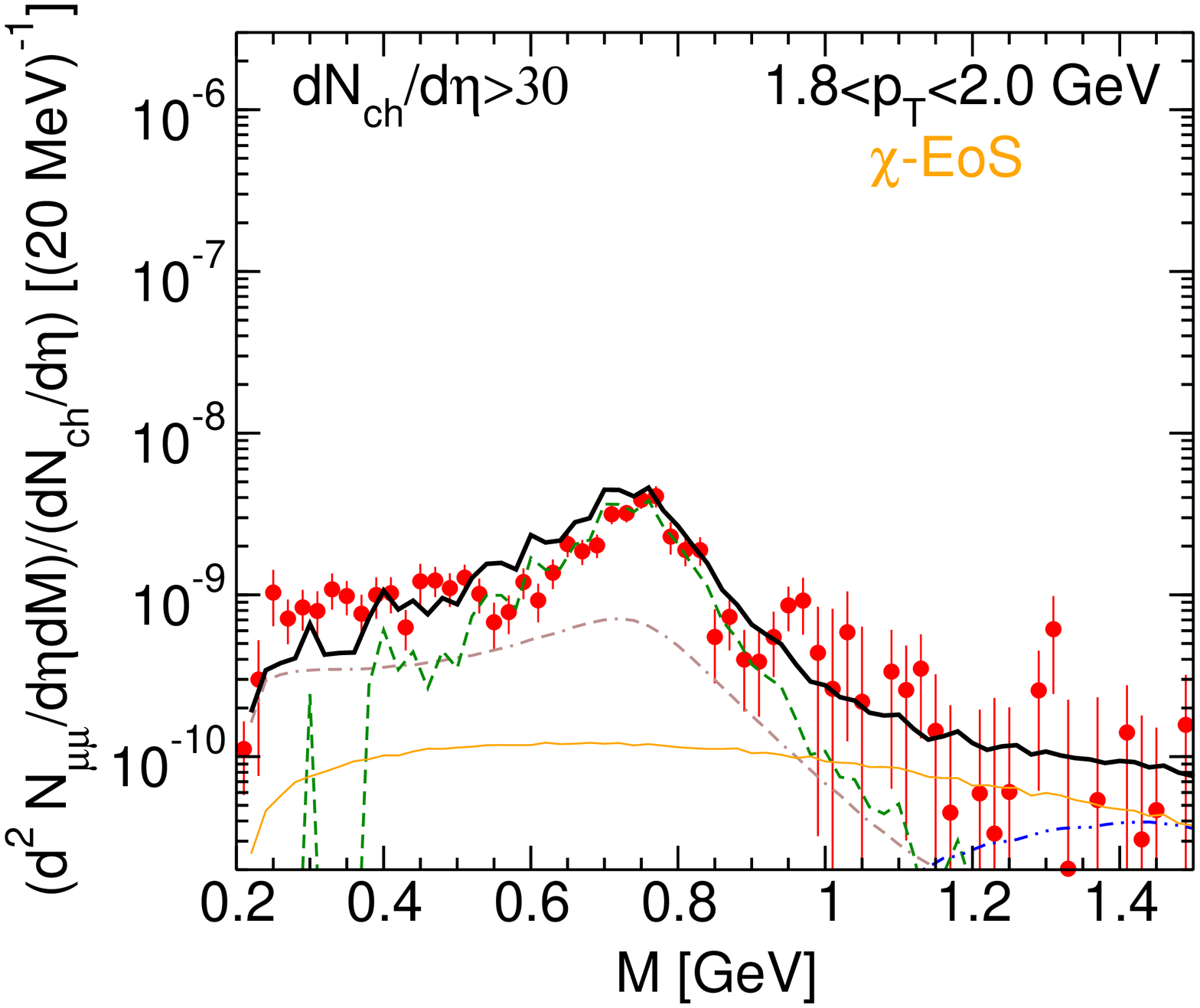}
\hspace{-1.4cm}
\includegraphics[width=.295\textwidth,angle=-90]{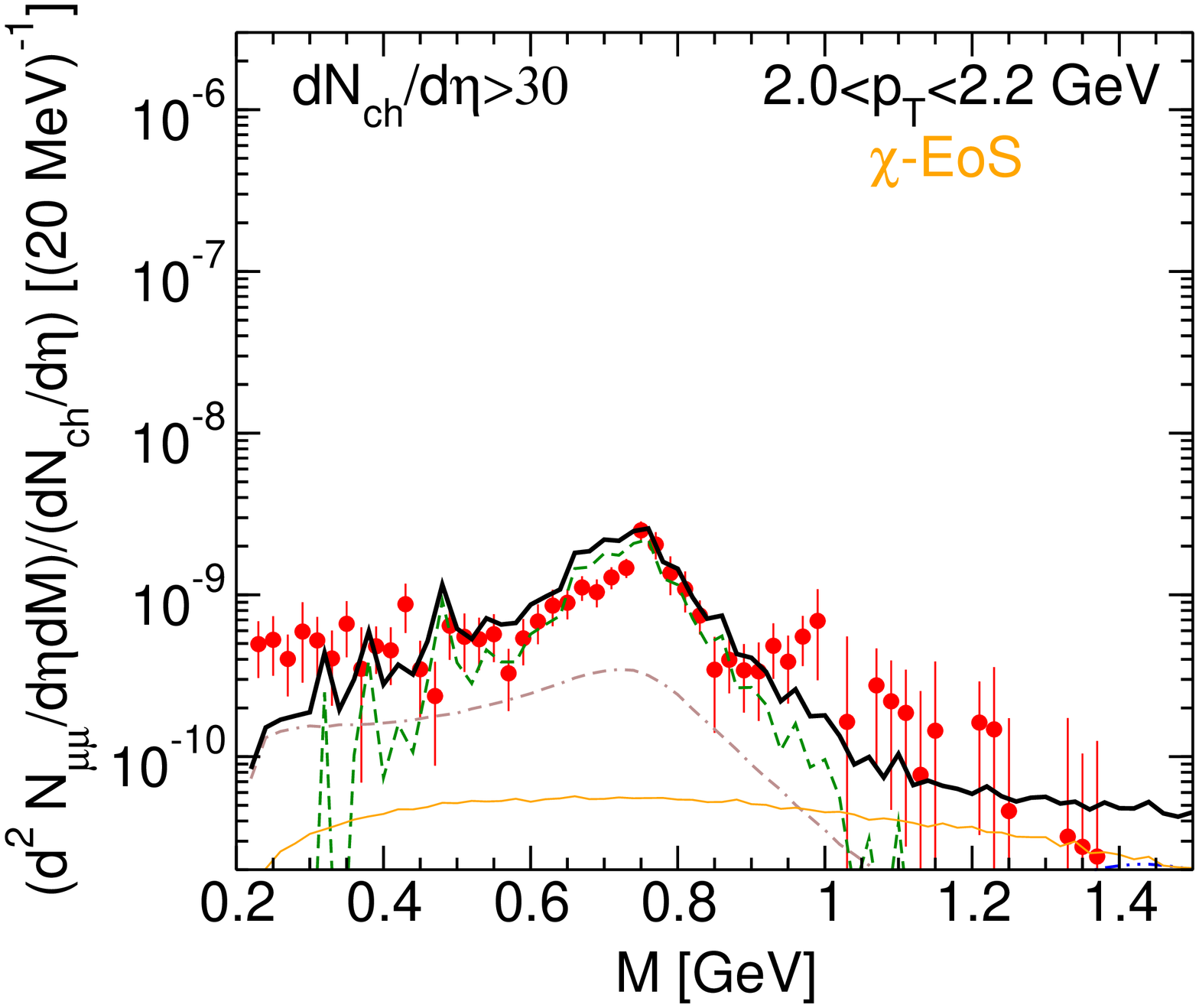}
\hspace{-1.4cm}
\includegraphics[width=.295\textwidth,angle=-90]{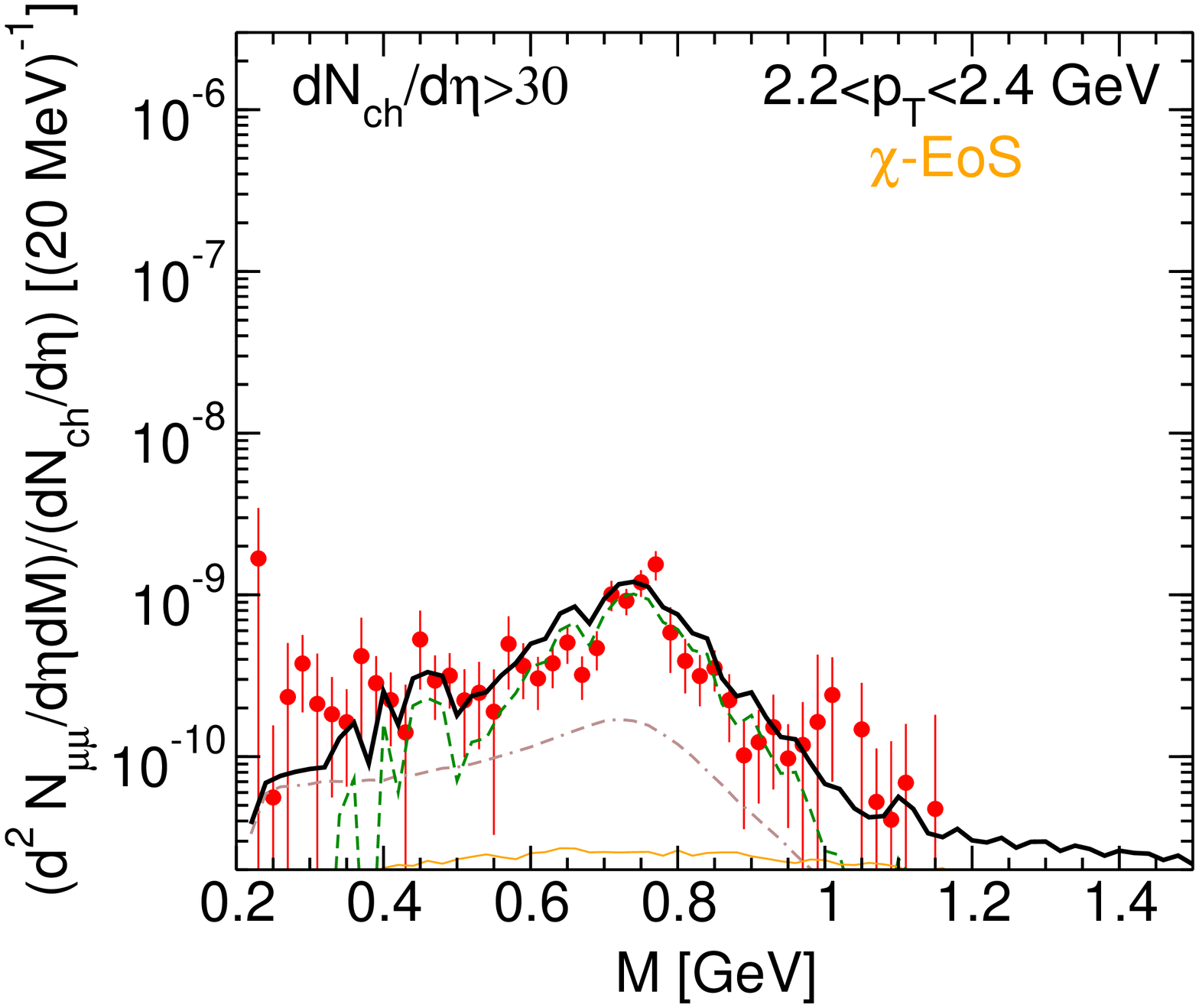}
\caption{(Color online) Acceptance-corrected invariant mass spectra of the 
excess dimuons in In-In collisions at 158$A$ GeV for various  bins of 
transverse pair momenta. 
The individual contributions arise from in-medium modified $\rho$ mesons 
(dotted-dashed line), $4\pi$ annihilation (double dotted-dashed line), 
quark-antiquark annihilation in the QGP (thin full line) and 
cascade $\rho$ mesons (dashed line). 
The sum of the various contributions is depicted by the thick full line.
Experimental data from Ref. \cite{Arnaldi:2008fw}. \label{fig1}}
\end{figure*}

In Fig. \ref{fig1} we show results for the invariant mass spectra of 
the excess dimuons in various slices in the transverse momentum of the 
dilepton pair $p_T$. The theoretical spectra are normalized to the corresponding average number of charged particles 
in an interval of one unit of rapidity around mid-rapidity 
\footnote{The rapidity window covered by the NA60 detector is about one unit 
in the forward direction. Here, we make use of the symmetry of the system 
and, in order to increase statistics, 
we apply the specular cut $|y|$$<$1 around mid-rapidity.}.
\begin{figure*}[t]
\includegraphics[width=.28\textwidth,angle=-90]{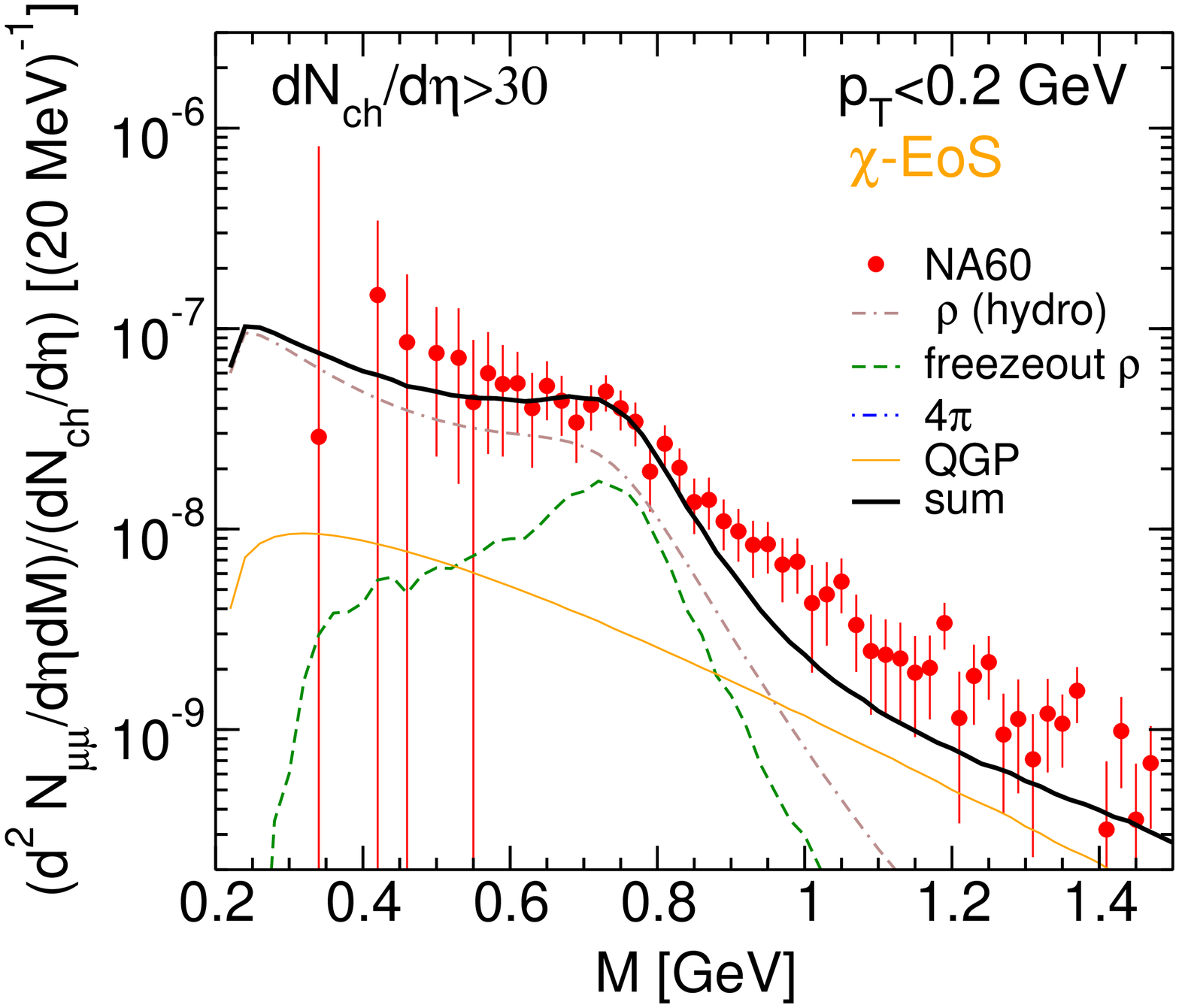}\hspace{-1.cm}
\includegraphics[width=.28\textwidth,angle=-90]{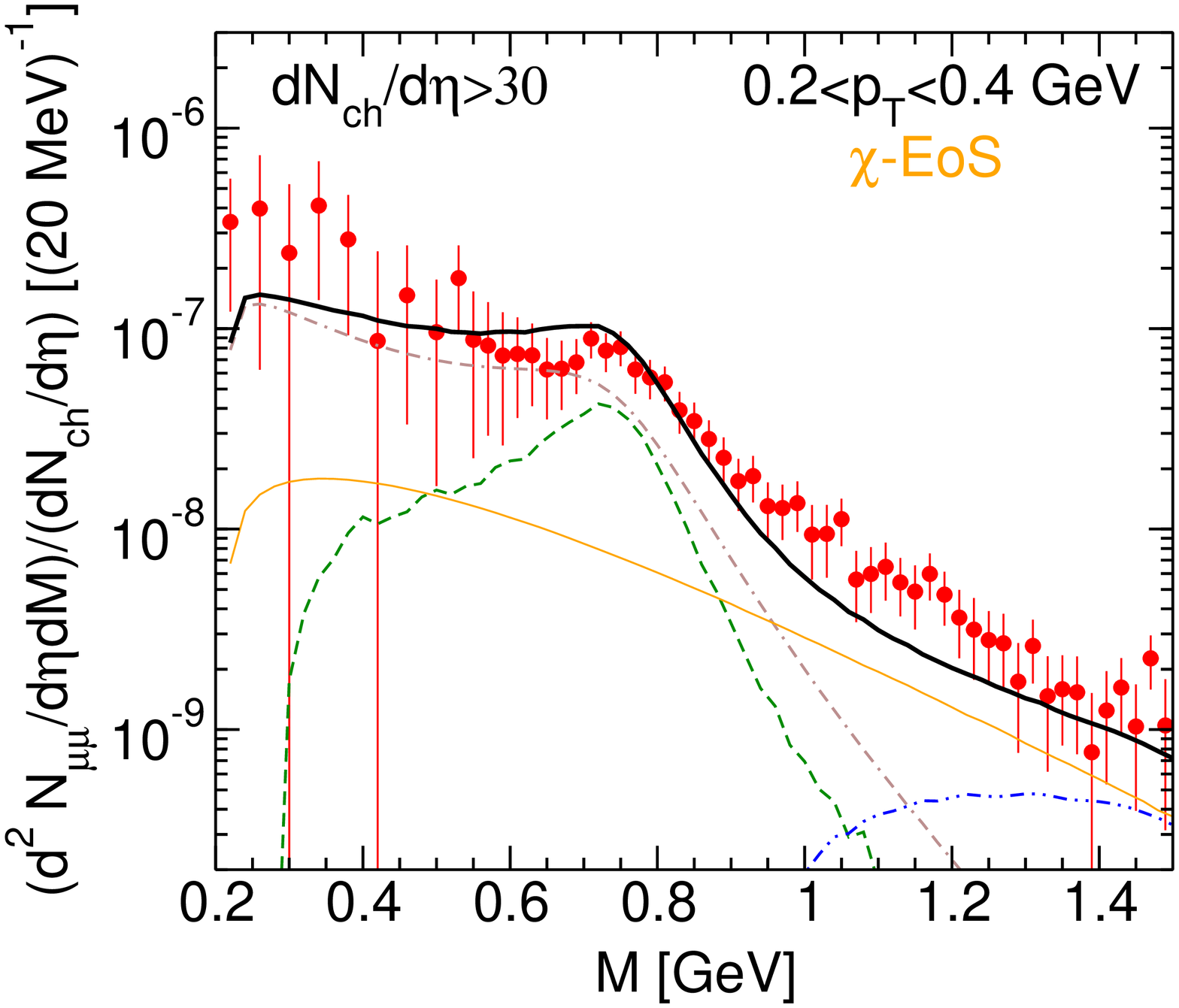}\hspace{-1.cm}
\includegraphics[width=.28\textwidth,angle=-90]{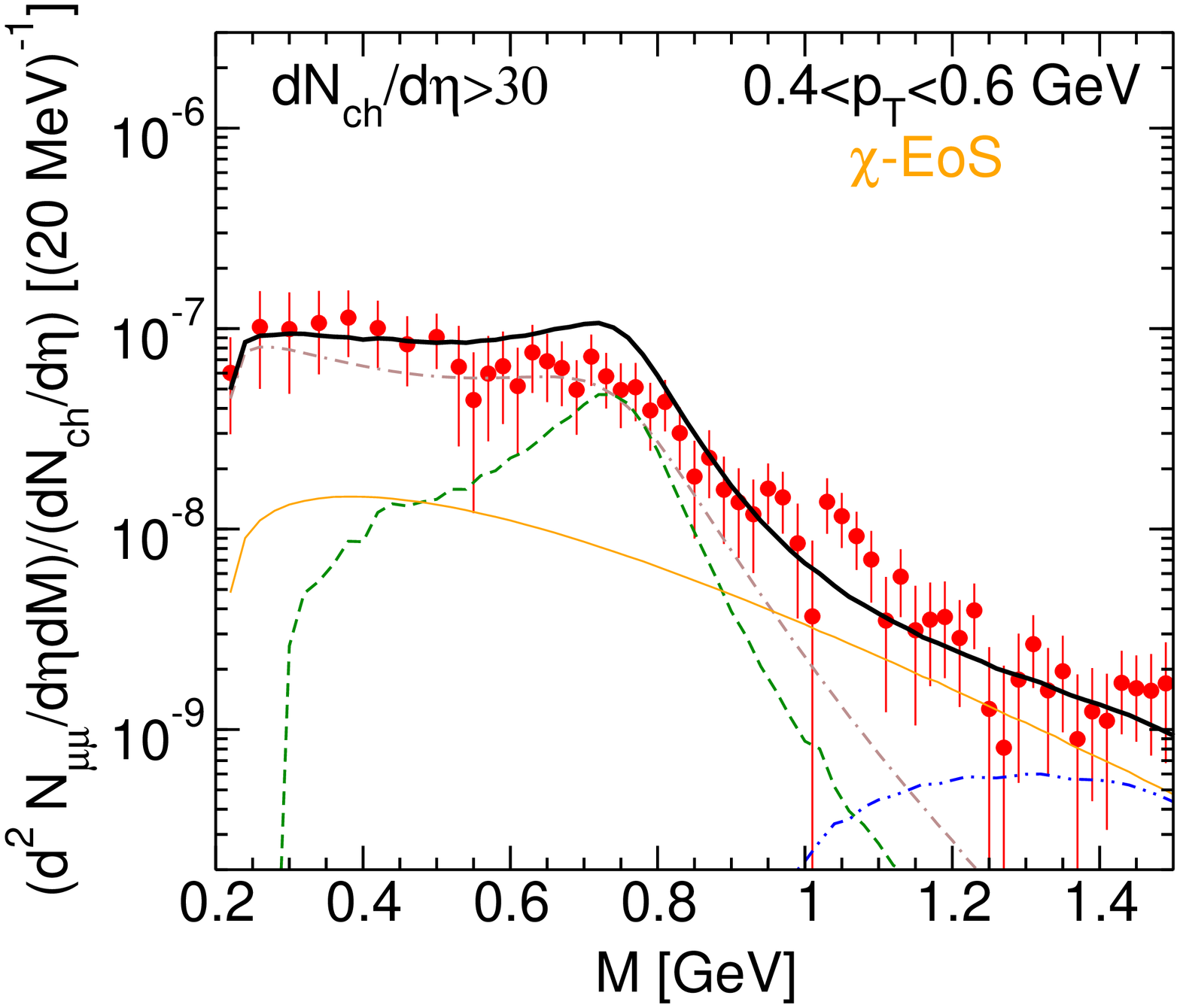}
\\
\noindent
\includegraphics[width=.28\textwidth,angle=-90]{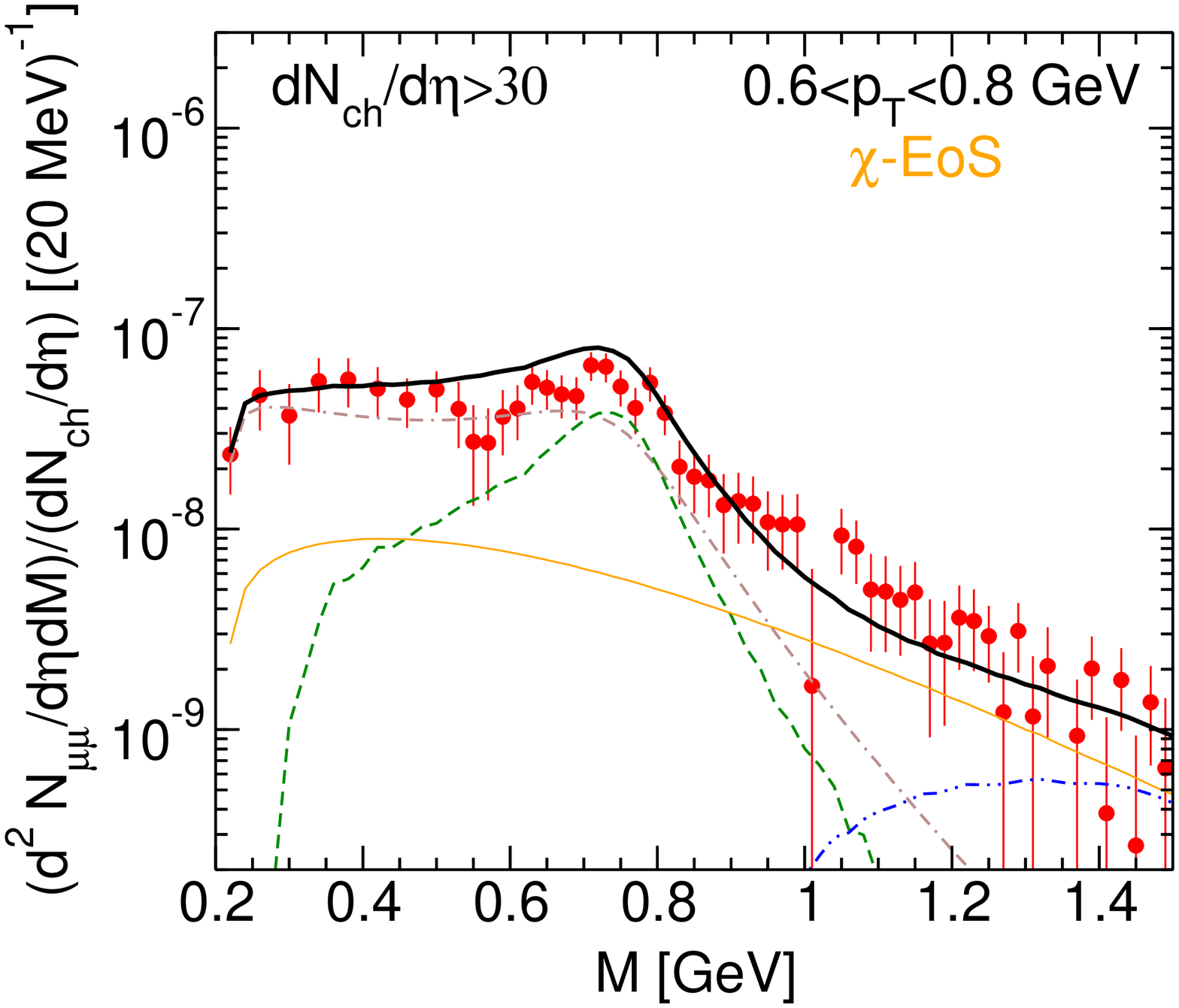}\hspace{-1.cm}
\includegraphics[width=.28\textwidth,angle=-90]{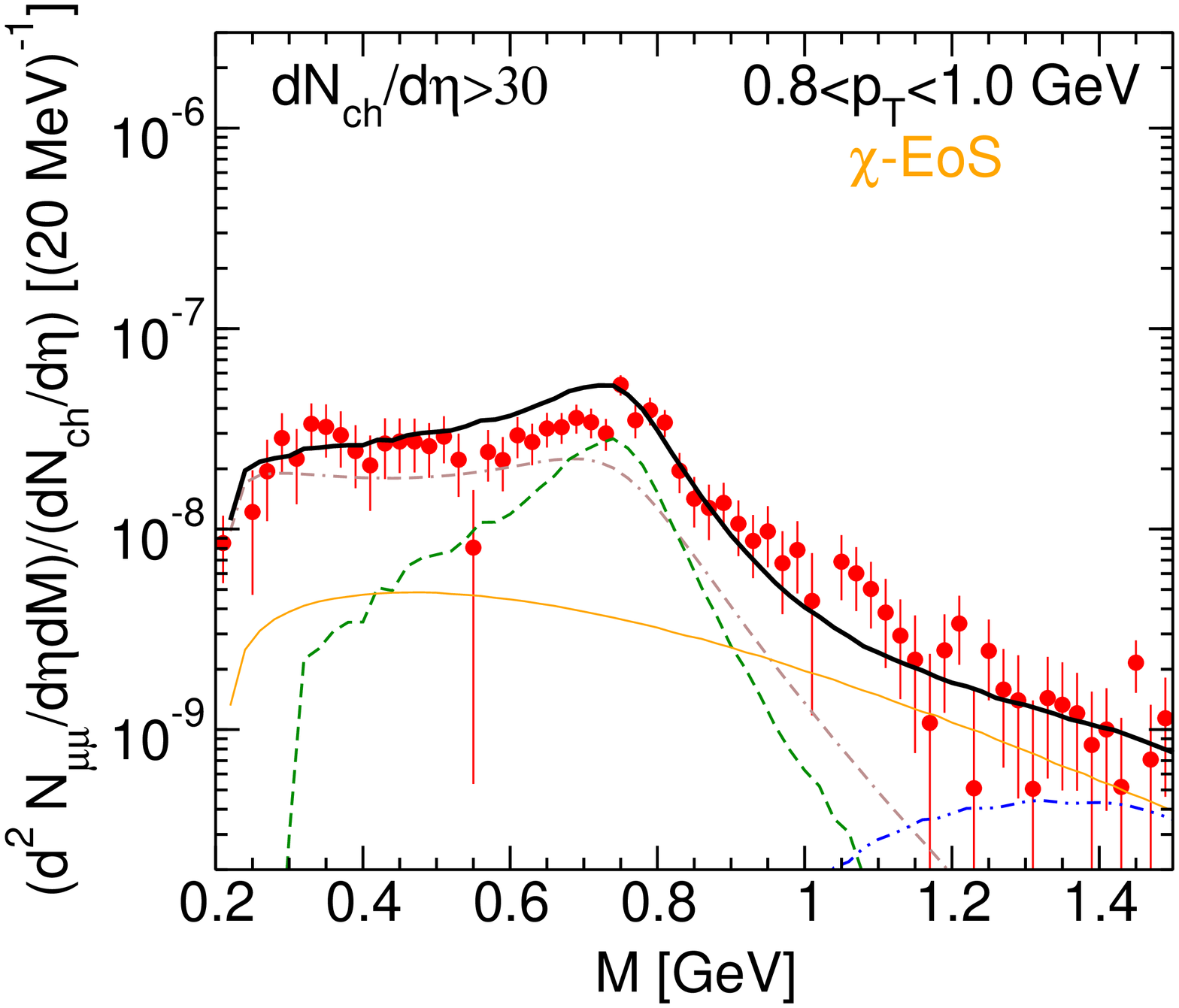}\hspace{-1.cm}
\includegraphics[width=.28\textwidth,angle=-90]{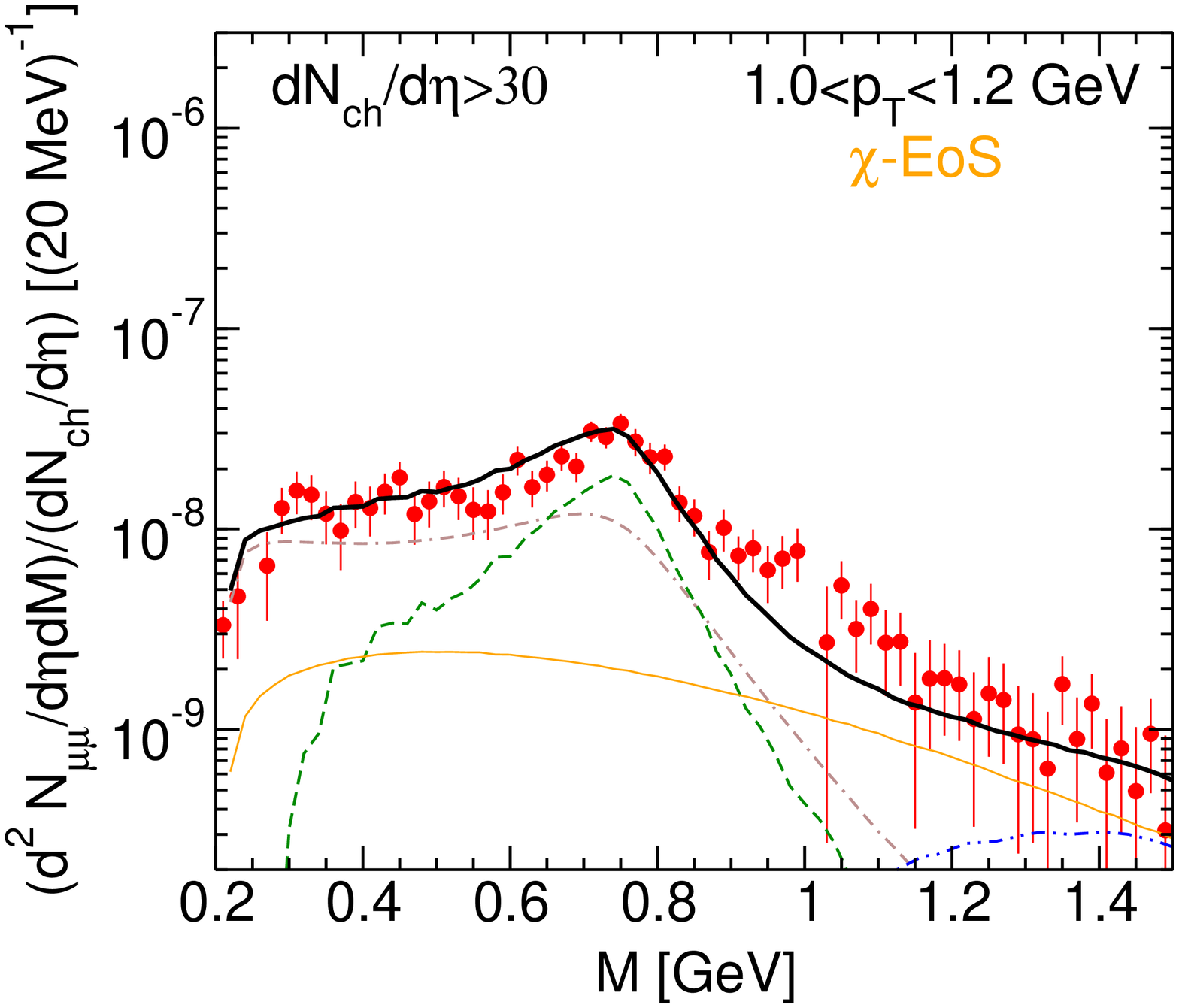}
\\
\noindent
\includegraphics[width=.28\textwidth,angle=-90]{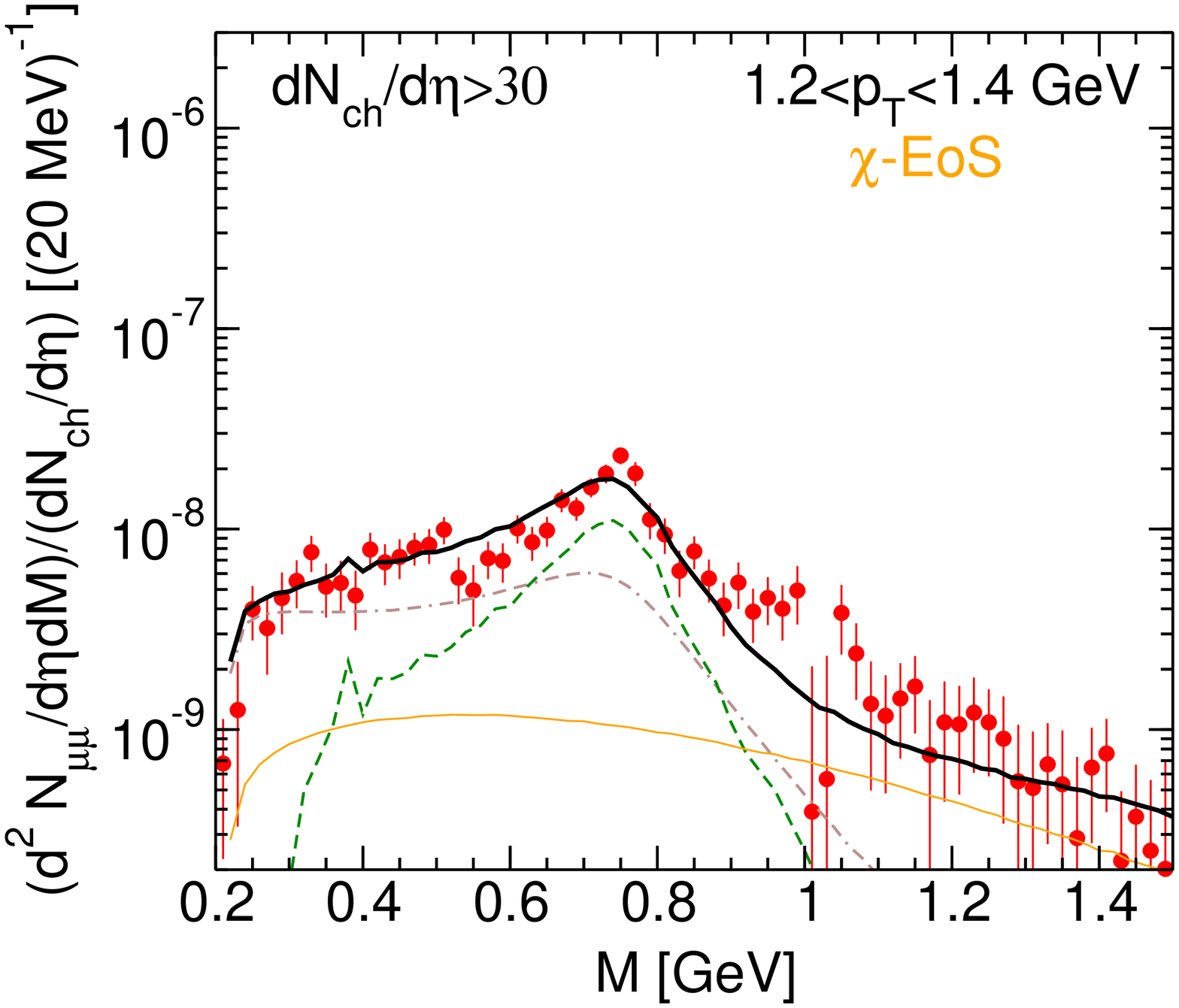}\hspace{-1.cm}
\includegraphics[width=.28\textwidth,angle=-90]{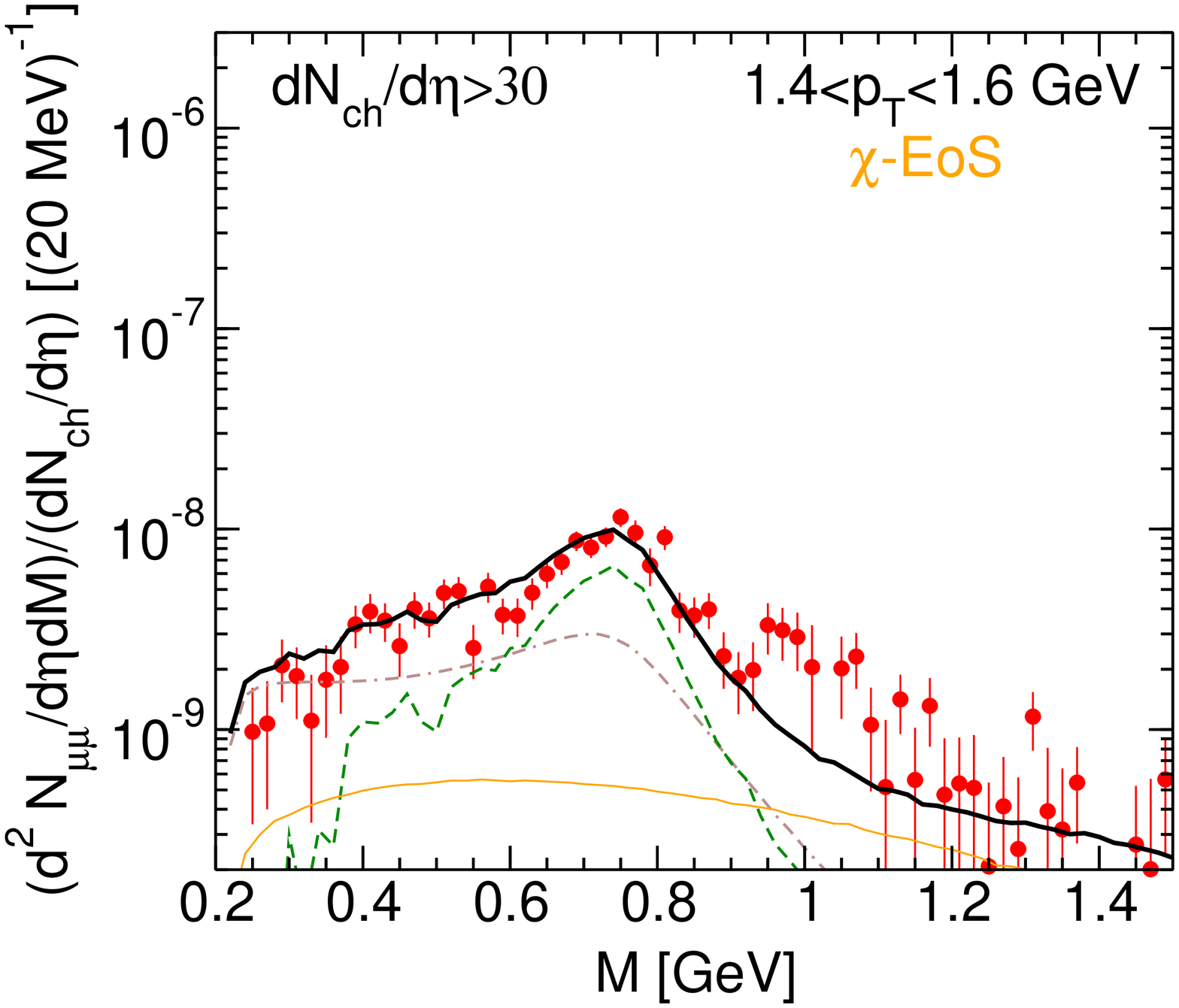}\hspace{-1.cm}
\includegraphics[width=.28\textwidth,angle=-90]{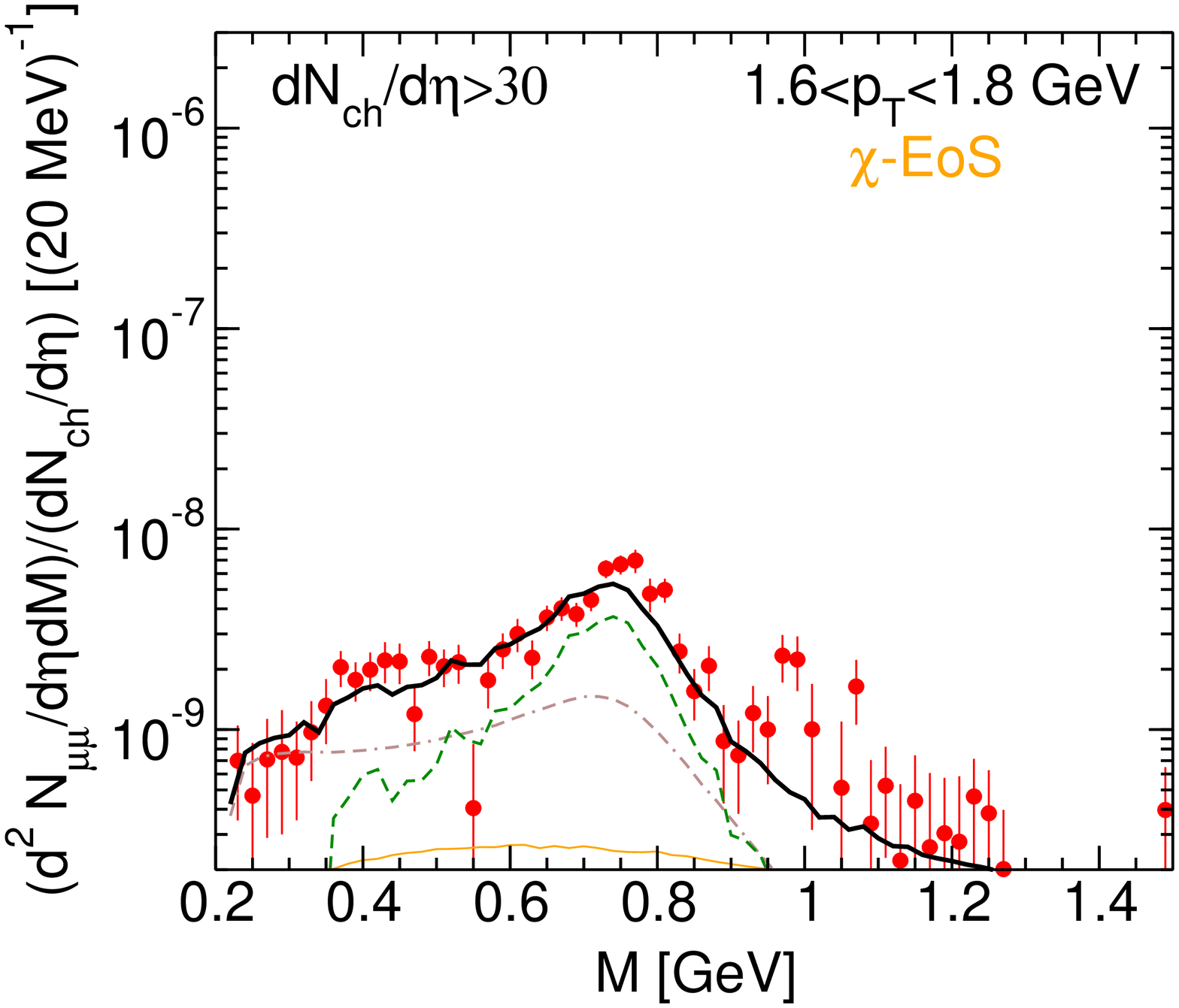}
\\
\noindent
\includegraphics[width=.28\textwidth,angle=-90]{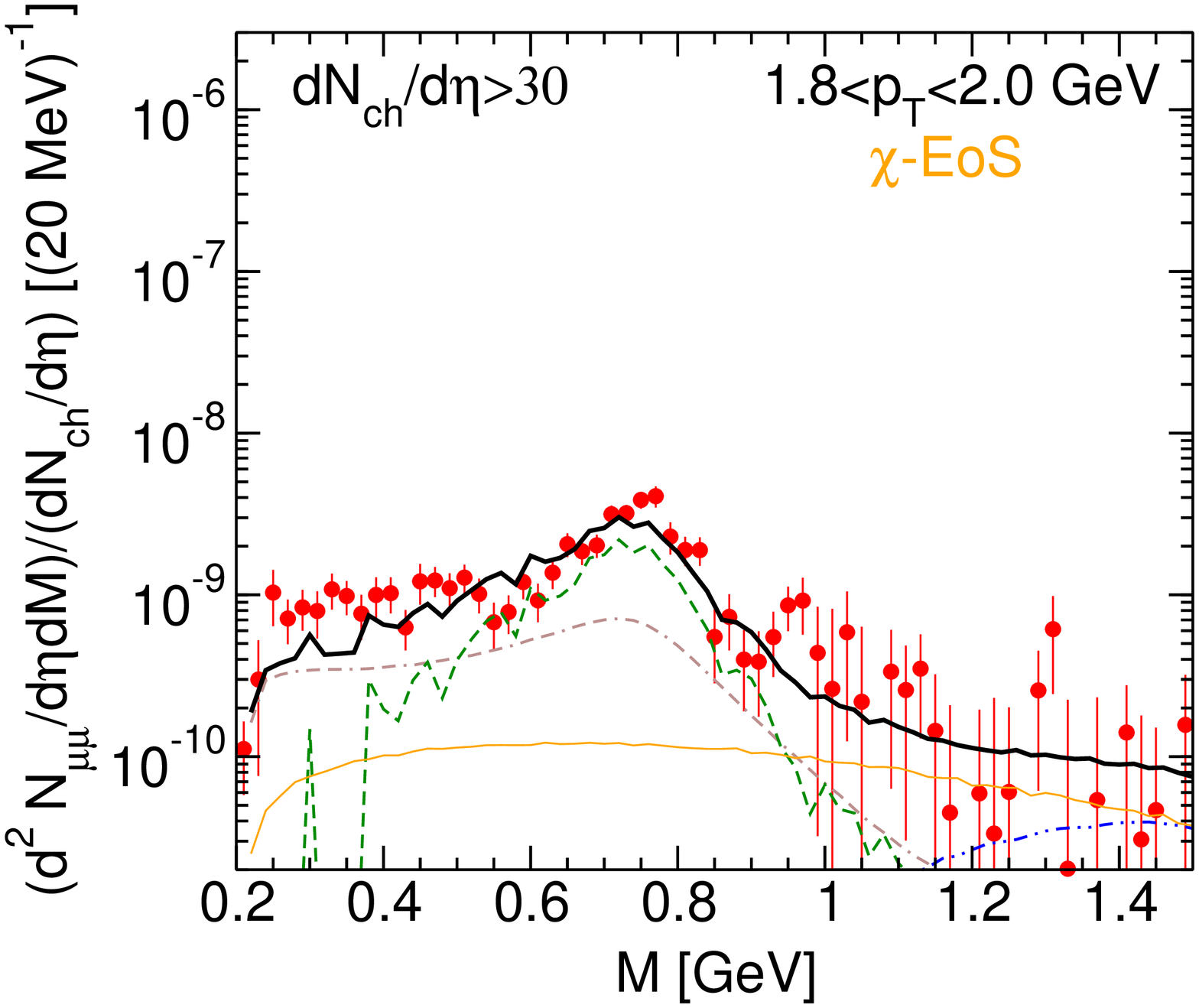}
\hspace{-1.cm}
\includegraphics[width=.28\textwidth,angle=-90]{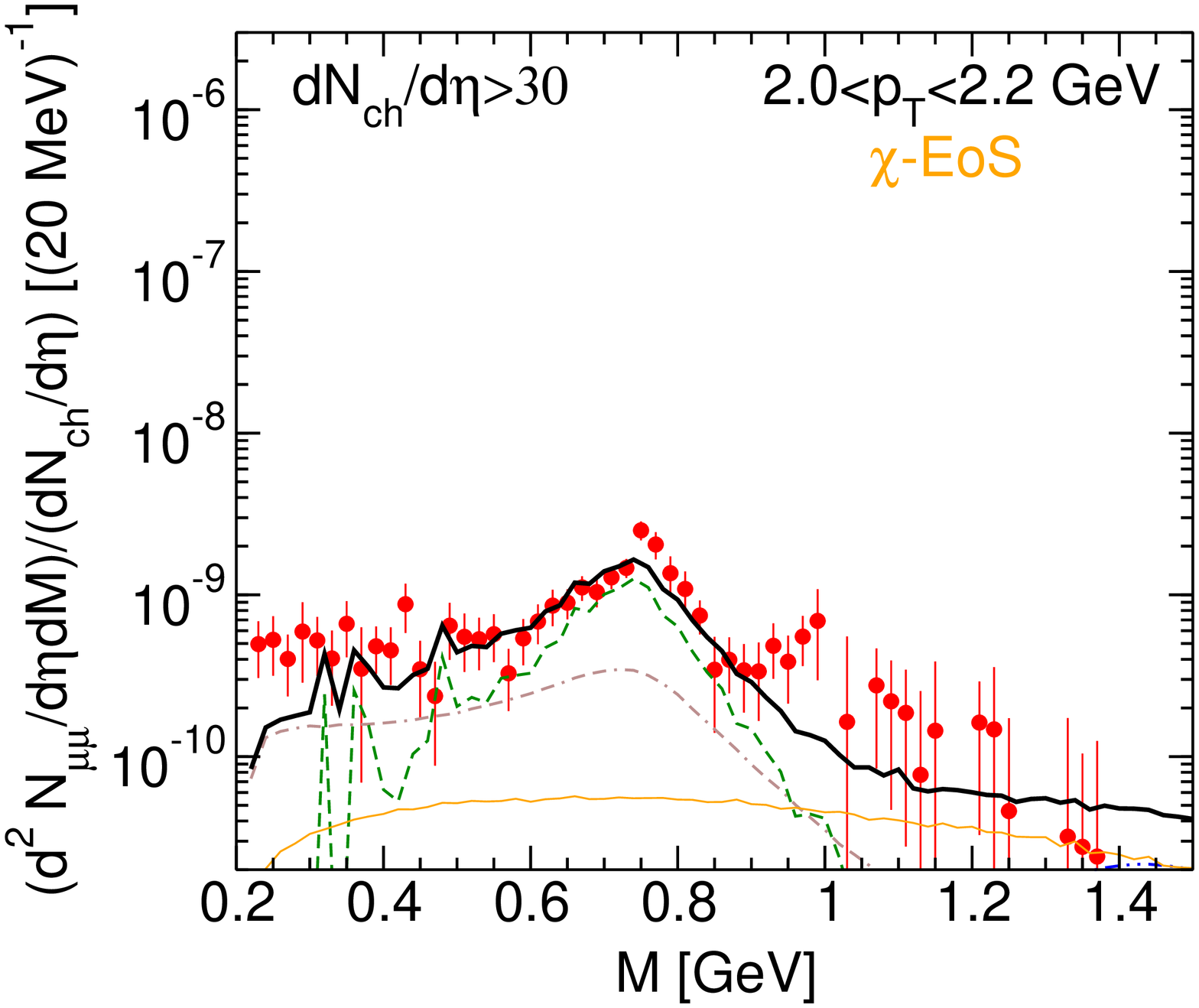}
\hspace{-1.cm}
\includegraphics[width=.28\textwidth,angle=-90]{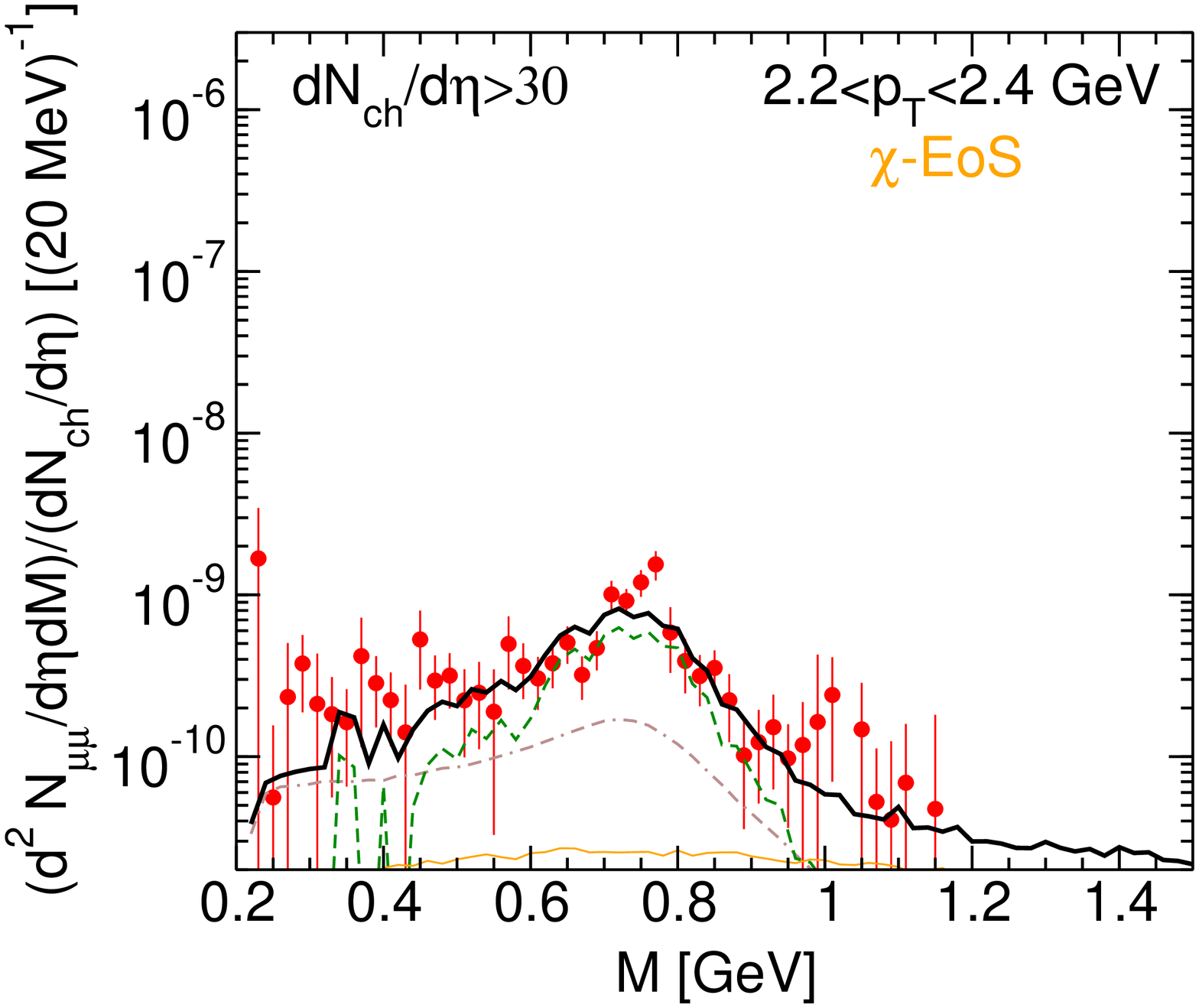}
\caption{Same as Fig. \ref{fig1}, but with the sudden freezeout approximation. \label{mspectra_sudcoop}}
\end{figure*}
In the invariant mass region $M$$<$0.5 GeV the spectra are 
dominated by the thermal radiation from the in-medium $\rho$ meson. 
The  $p_T$ scaling  exhibited compares nicely to the experimental observations. 

In the region around the $\rho$ meson peak, a net dominance of 
the cascade contribution is found. This result differs quite substantially 
from the one of 
previous theoretical work which adopted the sudden freezeout approximation. 
It is 
an intrinsic  feature of the hybrid model, due to the presence 
of a final cascade after the hydrodynamical evolution,  leading to a 
continuous and slow decoupling of the particles. In fact,
 the sum of the thermal and cascade contribution leads to an overestimation 
of the NA60 data in the peak region for $p_T$$<$1 GeV. 
With increasing $p_T$, however, the agreement with the experimental data improves. This aspect 
deserves further investigations and discussions but will be postponed 
to a separated dedicated section.
If we adopt the sudden freezeout approximation we obtain the result shown in 
Fig. \ref{mspectra_sudcoop}. One observes that the yield in the peak region is 
reduced in comparison to results obtained with a full cascade-like modelling of the 
time scale of the freezeout process and the global result is comparable 
to calculations by other groups (see Fig.4 of Ref. \cite{Arnaldi:2008fw}). 
The component denoted with freezeout $\rho$ and depicted by the dashed line 
corresponds to the dimuon yield from decays of primary $\rho$ mesons produced 
at the transition. 

In the intermediate mass region, 1$<$$M$$<$1.5 GeV, we find that emission from the QGP accounts for 
about half of the total radiation. The remaining half is filled by the 
considered hadronic sources. 
The $4\pi$ annihilation alone is comparable to the QGP emission only for 
$M$$>$1.4 GeV.
It is lower in the mass region 1$<$$M$$<$1.2 GeV, where, however the 
sum of the thermal and cascade $\rho$ is not yet completely negligible. 
With exception of the two lowest $p_T$ bins, where the data are slightly 
underestimated in the region 1$<$$M$$<$1.2 GeV, the description of the 
intermediate mass region is reasonable. 
As already highlighted in the very low mass region, the $p_T$ dependence 
of the thermal radiation in the intermediate mass region is compatible 
with experimental observations.
This latter point is noteworthy because in 
the three thermal approaches which have been used so far to compute dimuon 
spectra in comparison to NA60 data the transverse expansion is driven by two 
free parameters, namely 
the expansion velocity and the chosen radial profile 
(see Table 1 in Ref. \cite{Rapp:2009yu} for the specific values/profiles 
used by the three groups). 
In the present model, on the contrary, the transverse expansion 
during the hydrodynamical evolution is univocally determined by the pressure of 
the EoS and by the initial transverse profile generated by the preceding 
cascade. 
This represents an important improvement in the field of the dynamics of 
thermal dileptons. 

Finally, we find that the importance of the thermal contribution over the 
non-thermal one decreases with increasing $p_T$, in agreement with previous 
findings \cite{Renk:2006qr,Dusling:2007kh,Ruppert:2007cr,vanHees:2007th}.

\subsection{Transverse mass spectra}
\begin{figure}
\includegraphics[width=0.45\textwidth]{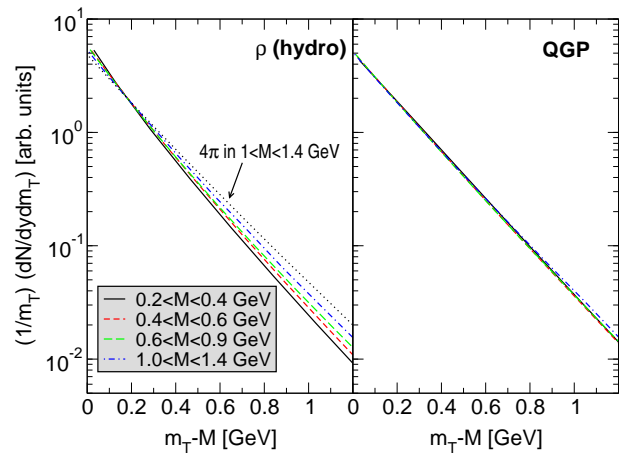}
\caption{(Color online) Transverse mass spectra of thermal dimuon emitted from in-medium $\rho$ (left panel) and QGP (right panel) in four mass windows. The dotted line depicts the emission from $4\pi$ annihilation processes in the mass window 
1$<$$M$$<$1.4 GeV. The spectra are arbitrarily normalized to increase readability.
\label{compare_slope}}
\end{figure}

We now reverse the analysis presented in the previous section and 
discuss transverse mass spectra of the dimuon excess for four different 
bins of invariant mass. 
Let us first focus on the thermal emission during the hydro phase and 
analyse the shape of the transverse mass spectra of two sources, 
the emission from QGP and from a hadronic source such as the in-medium $\rho$.
Dilepton radiation from these two contributions is shown 
Fig.~\ref{compare_slope}. 
For each mass bin the corresponding transverse mass spectrum 
has been separately scaled by an arbitrary factor 
to facilitate the visualization of the differences 
in the slopes of the four curves.
We observe that the  QGP emission (right panel) presents similar slopes for all 
four invariant mass bins. Such an independence from the mass clearly suggests 
early emission from a hot low flow source.  On the contrary, the radiation from the 
in-medium $\rho$ (left panel) exhibits a continuous 
hardening with increasing mass. 
Emission from a source with finite flow typically leads to such a mass 
ordering phenomenon. The same 
considerations explain why the $4\pi$ contribution, 
depicted by a dotted line in the left panel of Fig.~\ref{compare_slope}, 
is slightly harder than the 
thermal $\rho$ meson contribution in the last mass bin. Let us recall 
Fig.~\ref{fig1}  and focus on 
the invariant mass mass region 1$<$$M$$<$1.4 GeV: we note that  the $4\pi$ 
contribution is more copiously localised at masses higher than the in-medium 
$\rho$.

\begin{figure}
\includegraphics[width=.4\textwidth,angle=-90]{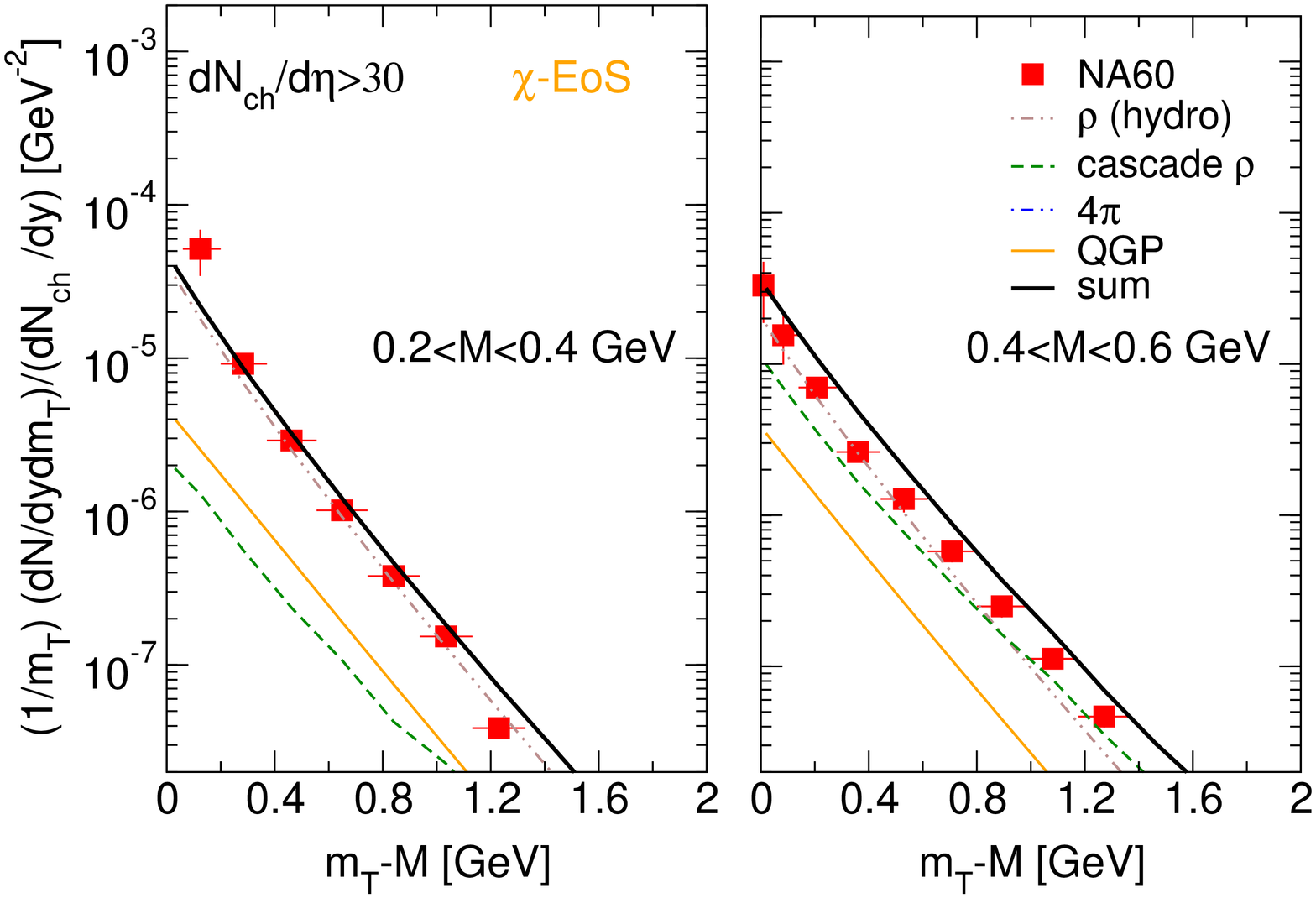}\\
\vspace{-1.cm}
\includegraphics[width=.4\textwidth,angle=-90]{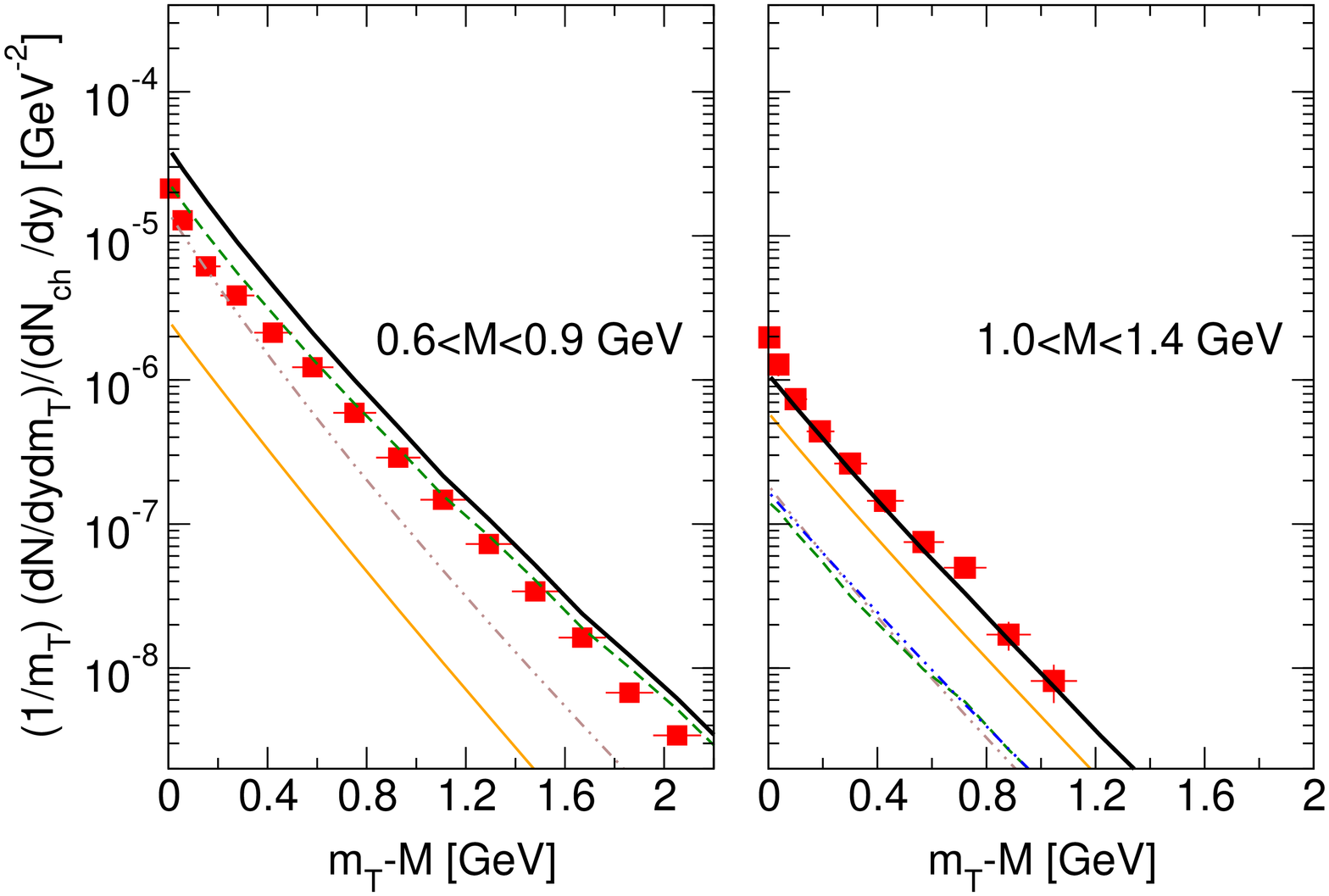}
\caption{(Color online) Acceptance-corrected transverse mass spectra of the excess dimuons 
in four mass windows. 
Experimental data from Ref. \cite{Arnaldi:2008fw}. \label{mt_spc}}
\end{figure}

Comparison of transverse mass spectra resulting from the present model and 
NA60 data is shown in Fig. \ref{mt_spc}.
The model fails to describe the peculiar rise at low $p_T$ ($m_T$$<$0.2 GeV) observed 
experimentally. At present, no consistent physical interpretation of the 
rise could be achieved \cite{Arnaldi:2008fw,Renk:2006qr}.
With exception of the lowest $p_T$ region, the model 
compares reasonably the experimental data in the lowest and highest mass bins. 
In the mass bin 0.4$<$$M$$<$0.6 GeV, the overestimation of the yield observed 
around the peak region starts to appear, but no strong deviation in the shape 
of the transverse mass spectra can be observed. This is not the case in the 
mass bin 0.6$<$$M$$<$0.9 GeV, where the most severe overestimation of the yield 
observed at low $p_T$ modify substantially the resulting shape of the 
transverse mass spectra 
and clear deviations with respected to the measured spectra are observed.

Finally, the lowest mass bin is dominated by the radiation from the 
in-medium vector meson, whereas in the other mass bins the resulting spectra 
are a composition of various sources, with exception of the high $p_T $ region of 
the 0.6$<$$M$$<$0.9 GeV bin, where the cascade contribution clearly dominates. 
Turning the discussion around, we can state that the lowest mass bin can be 
considered as barometer of the radiation emitted by the in-medium $\rho$. 
The reasonable agreement with the experimental observations suggests that the latter does 
indeed follow a dynamical path as expected from hydrodynamical models. 

\begin{figure}
\includegraphics[width=.4\textwidth,angle=-90]{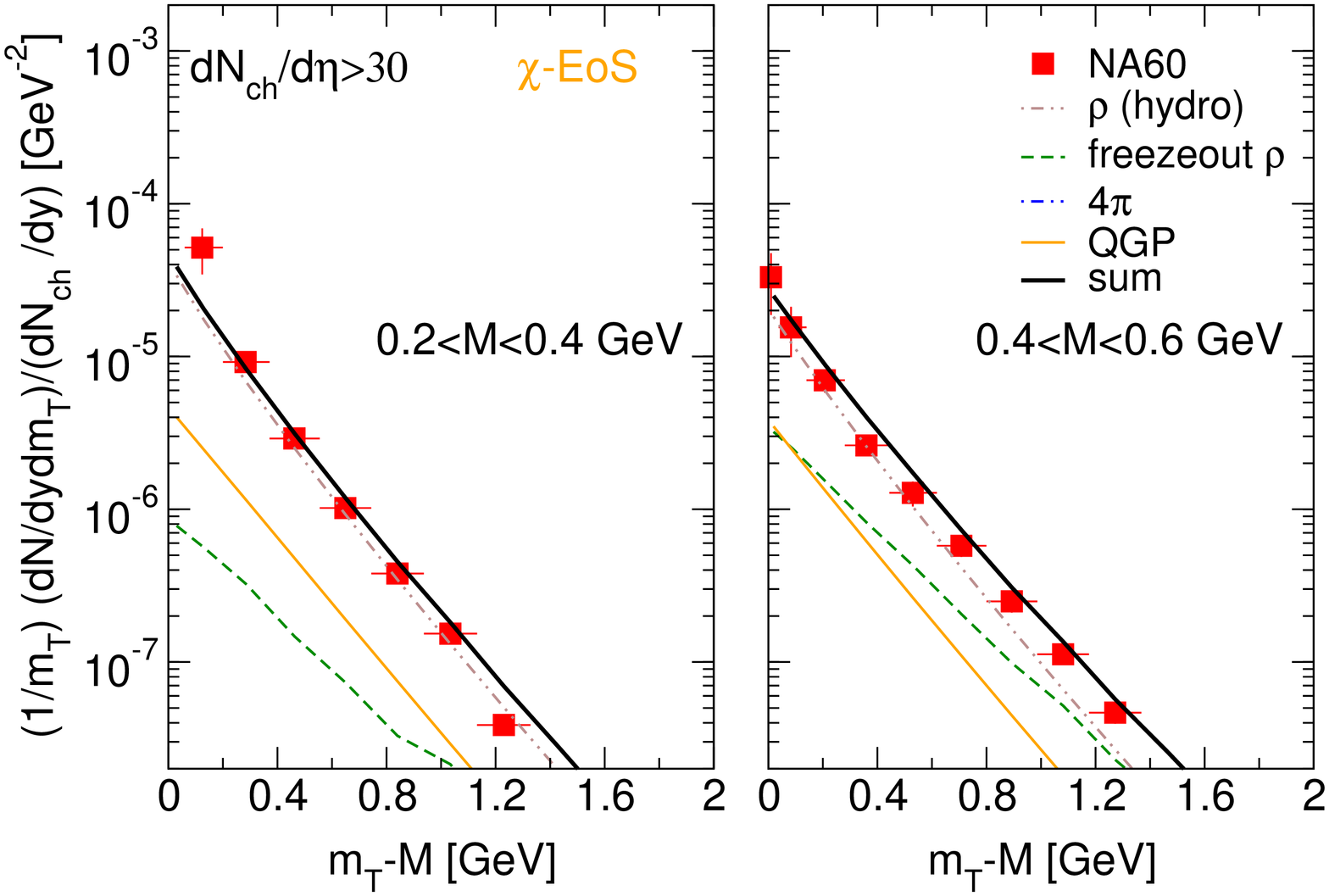}\\
\vspace{-1.cm}
\includegraphics[width=.4\textwidth,angle=-90]{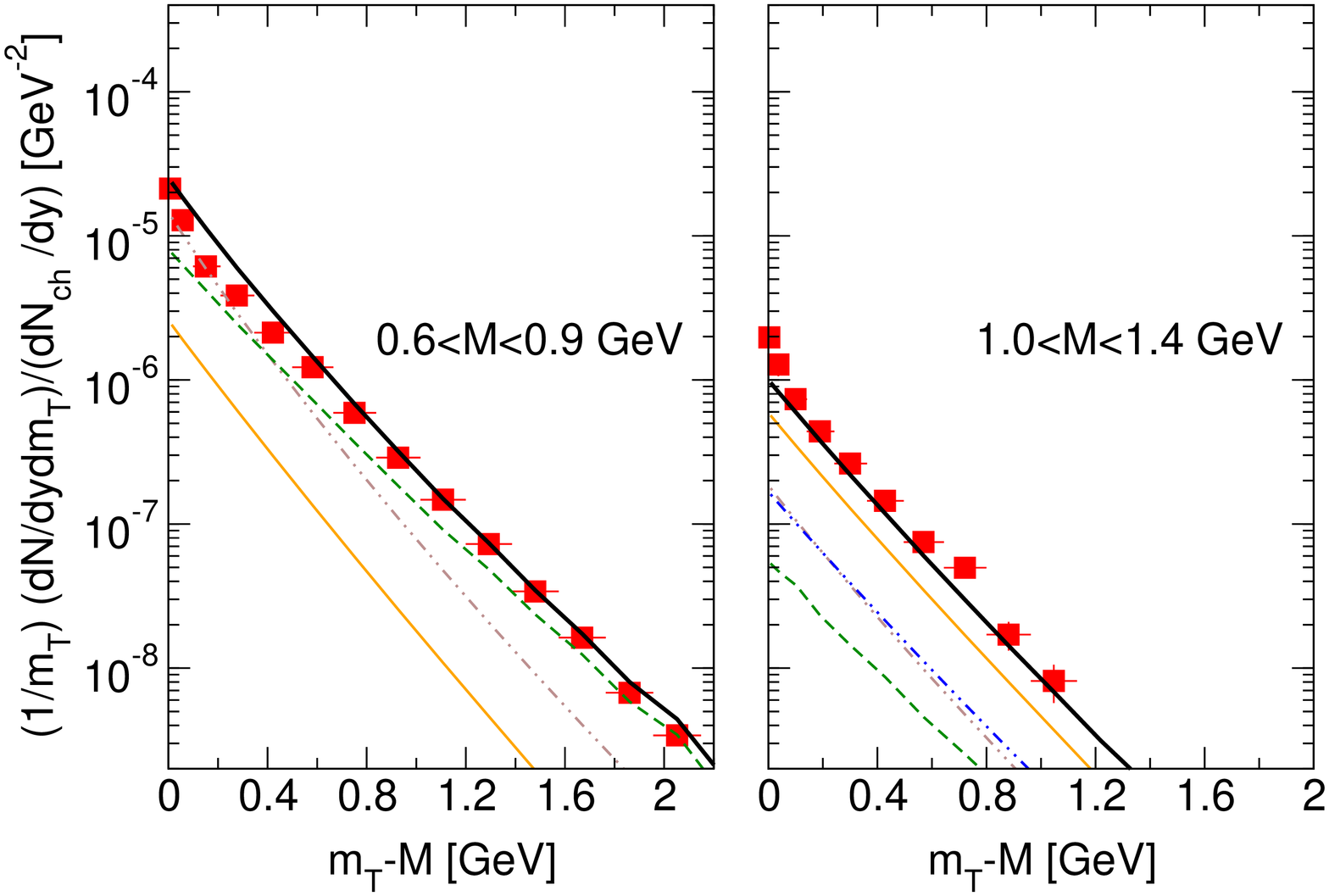}
\caption{(Color online) Same as Fig. \ref{mt_spc}, but with the sudden freezeout 
approximation. \label{mt_spc_sudcoop}}
\end{figure}

If we adopt the sudden freezeout approximation we obtain the result shown in 
Fig.~\ref{mt_spc_sudcoop}. This is practically identical to the previous one in the lowest and highest mass bins, whereas major differences can be observed in 
the two intermediate mass bins and are caused by the reduction 
of the non-thermal emission.


\begin{figure*}
\includegraphics[width=0.9\textwidth]{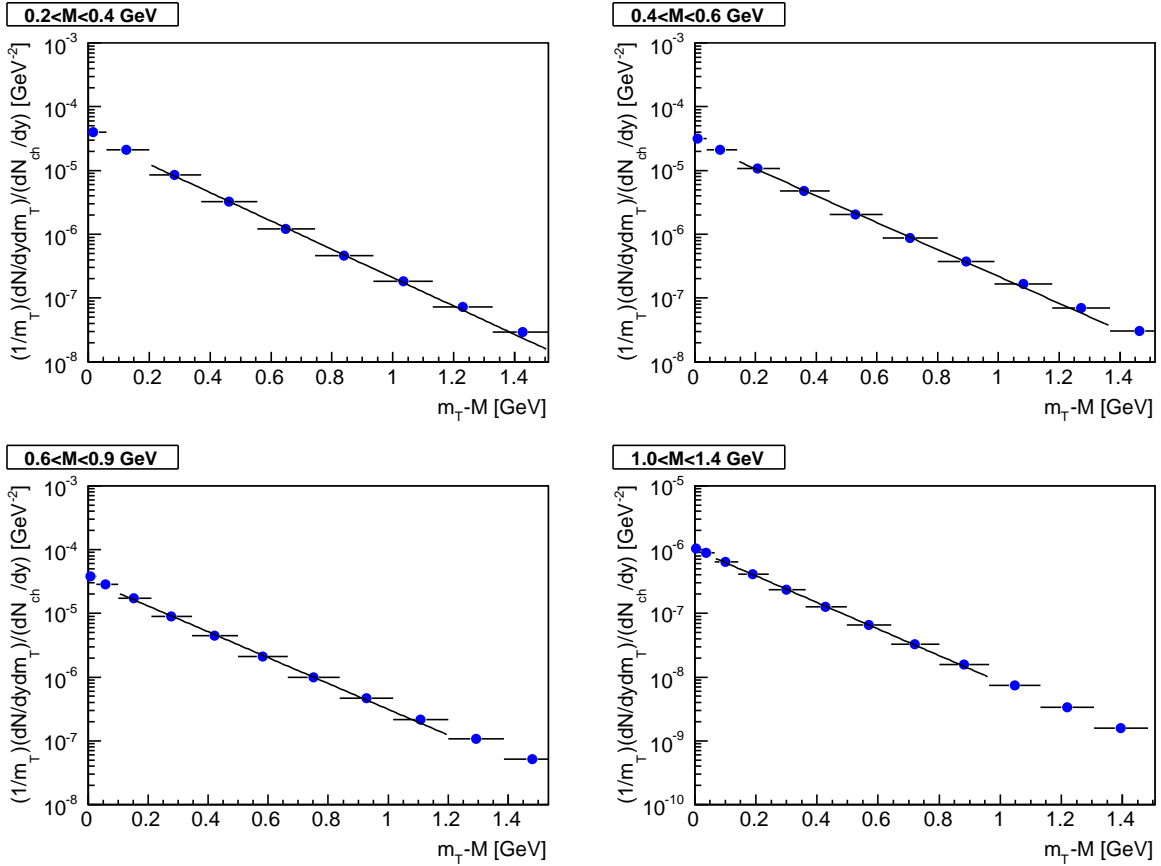}
\caption{(Color online) Exponential fit of the transverse mass spectra of the excess dimuons in four mass windows. The fit range is restricted to the transverse momentum 
interval 0.4$<$$p_T$$<$1.8 GeV (Note that the fit range in $m_T-M$ is different). \label{fit}}
\end{figure*}

\begin{figure}
\includegraphics[width=.45\textwidth]{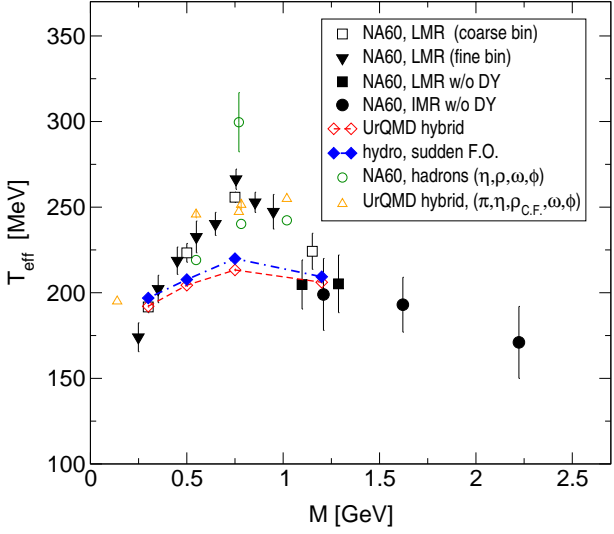}
\caption{(Color online) Inverse slope parameter $T_{\rm eff}$ vs. dimuon invariant mass. Experimental data (full triangles, full and open squares, full circles) are from Ref.~\cite{Arnaldi:2007ru,Arnaldi:2008fw}. The values of $T_{\rm eff}$ extracted by the fit of the calculated transverse mass spectra (shown in Fig.~\ref{mt_spc}) are indicated by open diamonds. The full diamonds depict the  values of $T_{\rm eff}$ extracted by the fit of the transverse mass spectra obtained using the sudden-freeze-out approximation (shown in Fig.~\ref{mt_spc_sudcoop}). The open circles depict hadron
data obtained at NA60 as a by-product of the cocktail subtraction procedure ($\eta$, $\omega$ and $\phi$) and from a decomposition into peak and continuum of the $\rho$-like window (see Ref.~\cite{Arnaldi:2008fw} for details). The corresponding hybrid model results are indicated by open triangles. \label{teff}}
\end{figure}

To complete our analysis of the transverse mass spectra of the dimuon excess, 
we follow the procedure adopted by the NA60
collaboration and perform a quantitative analysis of effective slope
parameters. The
measured spectra have been divided into several invariant-mass
bins, in each of which the data have been fitted with the function 
$(1/m_T)dN/dm_T\propto \exp(-m_T/T_{\rm eff})$, 
where the effective temperature parameter 
$T_{\rm eff}$ is the inverse slope of the distribution. 
The fit range was taken as 0.4$<$$p_T$$<$1.8 GeV identically 
to the NA60 fit range\cite{Arnaldi:2008fw}.
We apply this procedure to the results of our calculation.
The resulting exponential fits are shown in Fig.~\ref{fit}.
In Fig.~\ref{teff} the inverse slope parameters extracted from the 
fit procedure are plotted vs. the dimuon mass, as 
done by the experimental collaboration. We observe, that the model is able 
to qualitatively describe the rise of $T_{\rm eff}$ with mass up to the pole position of the $\rho$ followed by a drop in the intermediate mass region observed experimentally. However, at the quantitative level, the extracted values of  
$T_{\rm eff}$ reproduce the experimental ones only in the first and last mass 
bin, whereas underestimate them in the second and third mass bin. 
One should note that a variation in the fit range, taking e.g. the fit range 
chosen in Ref.~\cite{vanHees:2007th}, would increase the 
temperature by 8\% in the first three mass bins and 3\% in the latter, 
so that the fall down would persist. However,  such a fit range would not 
be consistent with the experimental procedure.


At present, we can only speculate on the origin of the discrepancies  in the values of the effective temperatures extracted in the mass region 0.4$<$$M$$<$0.9 GeV. The excitation function of the mean transverse mass of various hadrons has been examined in previous works~\cite{Petersen:2008dd,Petersen:2009mz,Steinheimer:2009nn} where it was shown that, whereas the hybrid approach reasonably reproduces the general behaviour of the mean $m_T$ as a function of energy for various particle species, slightly higher values than the experimental data were observed for pions, protons and negatively charged kaons produced in nucleus-nucleus collisions at the top SPS energies. 
Similar considerations can be drawn from the comparison of the inverse slope parameters extracted for the $\eta$, $\omega$ and $\phi$ mesons by the NA60 Collaboration from the analysis of the cocktail (Fig.~\ref{teff}, open circles) with the respective values obtained within the present model (Fig.~\ref{teff}, open triangles). For completeness, the model result for the effective temperature of negatively charged pions~\footnote{For the pions, the fit has been performed in the range 0$<$$m_T-m_0$$<$0.7 GeV; for the other hadrons, the same fit range as used for the excess, i.e. 0.4$<$$p_T$$<$1.8 GeV, has been used.} has been inserted in Fig.~\ref{teff} as well. We note that the effective temperatures of the three cocktail hadrons are slightly overestimated by the hybrid approach. Thus, a more copious radial flow than the one resulting from the present model can be hardly conciliated with the overall hadron results at the same beam energy.  

The model predicts for the $\phi$ an only slightly higher value of $T_{eff}$ than for the $\omega$, in qualitative agreement with the experimental observations, but on the same time returns comparable values for the $\omega$ and the $\rho$ mesons, where with ’’$\rho$’’ here we intend the contribution to the total emission in the mass bin 0.6$<$$M$$<$0.9 GeV from freeze-out $\rho$ as resulting when adopting the sudden freeze-out approximation in the calculation. On the contrary, the inverse slope parameter 
$T_{eff}$ extracted by the NA60 data for the peak region, isolated from the continuum in the mass spectrum by a side-window subtraction method and interpreted as the freeze-out $\rho$~\cite{Arnaldi:2008fw}, is about 50 MeV higher than that extracted for the $\omega$ mesons, of similar mass, and above the one measured for protons in Pb-Pb collisions, which are also strongly coupled to pions \cite{Arnaldi:2010zz}. Such a behaviour can be hardly obtained in models in which decoupling from the hydrodynamical description happens simultaneously for all particles species, as the present. 
Nevertheless, as will be shown in the next section, a close analysis of the cascade stage that follows the hydrodynamical evolution does suggest that after the transition the $\rho$ meson is still coupled to the medium.  Within the present model, however, essentially no additional radial flow is developed in this late stage since the cascade description is characterized by softer transverse dynamics than the hydrodynamics one, so that the maximum coupling to the flow is typically reached at the end of the hydrodynamical evolution. This may suggest that the stage during which the $\rho$ emerges from the hydrodynamical flow, with corresponding modifications of its spectral function, is too roughly described by the approximation that, instantaneously and within a layer, one may transit from an in-medium modified $\rho$ coupled to the fluid to a vacuum-like $\rho$ interacting according to a cascade description. Presumably, residual in-medium modifications are still present during the time-span of decoupling and the cascade stage. Escape probability arguments, as well as the analysis of the cascade stage, suggest that the ability to decouple is momentum dependent since fast particles escape earlier than slow ones. To take all these effects into account properly is not trivial and beyond of the aims of the present  work. Generally such models can be constructed as e.g. done by Grassi et al.~\cite{Grassi:1994nf}.
Indeed, an indirect evidence for the presence of in-medium modifications of the low $p_T$ cocktail $\omega$'s has been observed by the NA60 collaboration in the form of disappearance of the yield in the low $m_T$ region with respect to a reference exponential fit line~\cite{Arnaldi:2008fw}. If one assumes a similar behaviour for the $\rho$ mesons, then a depletion at low $m_T$ of the freeze-out contribution to the transverse mass spectra can be expected. Whether this can effectively result in larger values of the inverse slope parameter of the whole excess depends substantially on the relative importance of the freeze-out and thermal contribution in the low $m_T$ region.

Another effect that may harden the transverse dynamics is the inclusion of viscosity in the hydrodynamical equations, as schematically shown in Refs.~\cite{Dusling:2008xj,Song:2010fk}. However, the viscous effects induce a hardening of the transverse spectra not only of dileptons but also of pions~\cite{Song:2010fk}, therefore a consistent interpretation of the transverse spectra of hadrons and dileptons at top SPS might be challenging even for viscous hydrodynamics.

\section{Decomposition of the cascade contribution 
and discussion \label{decomp}}
In this section we want to have a closer look to the various components 
that enter the contribution of the cascade stage to the dimuon spectra. 
The total cascade emission consists of: (i) the emission from the stage that 
precedes the hydrodynamical evolution (denoted as ``pre-hydro''); (ii) the 
dimuon emission from the primary $\rho$'s produced at the point of transition 
between the hydrodynamic and the cascade evolution 
via the Cooper-Frye equation (denoted as ``Cooper''); 
(iii) the emission from $\rho$ mesons (re)generated from interactions 
occurring during the late cascade stage that follows the 
hydrodynamics evolution (denoted as ``regenerated''). 
\begin{figure}
\includegraphics[width=.45\textwidth]{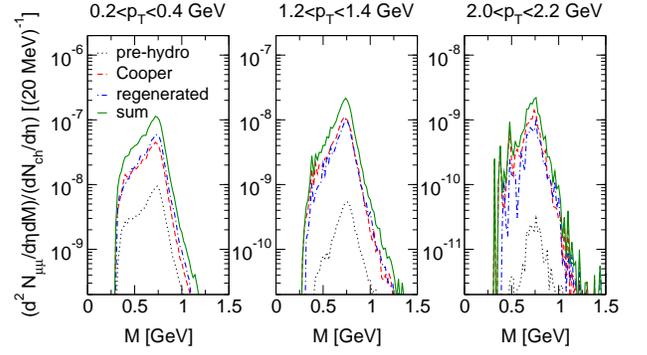}
\caption{(Color online) Decomposition of the cascade $\rho^0$$\rightarrow$$\mu^+\mu^-$
emission in: (i) emission from the stage that precedes 
the hydrodynamical evolution (dotted line); 
(ii) emission from $\rho$ mesons which are merged
into the cascade at the transition point via the Cooper-
Frye equation and emit in the cascade stage of the evolution
(dashed line); (iii) emission from $\rho$ mesons which are produced
and emit in the cascade stage (dotted-dashed line). The full
line represents the total cascade emission.
\label{cascade}}
\end{figure}
\begin{figure}
\includegraphics[width=.35\textwidth]{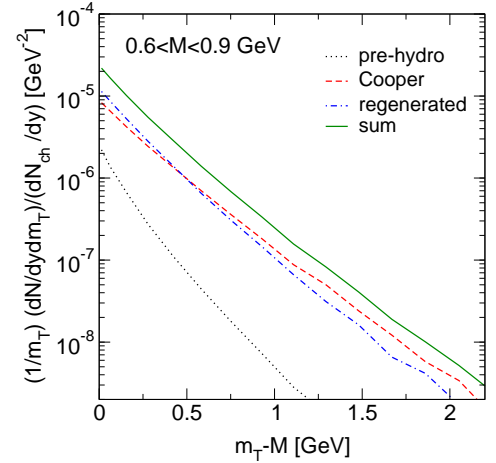}
\caption{(Color online) Decomposition of the cascade $\rho^0$$\rightarrow$$\mu^+\mu^-$
emission as in Fig. \ref{cascade}, but for transverse mass spectra.
\label{cascade_pt}}
\end{figure}

In Fig. \ref{cascade} we show the three contributions for three 
different bins on transverse momentum, a low (left), 
an intermediate (center) and a high one (right). 
The emission from the stage that 
precedes the hydrodynamical evolution (dotted line) is negligible in comparison 
to the others. The contribution to dimuon emission from $\rho$ mesons 
generated in the cascade stage (dotted-dashed line) is in general comparable 
to the one from 
$\rho$ mesons created at the transition (dashed line), with 
its relative importance decreasing 
for increasing transverse momentum. This latter pattern is due to the 
generally more moderate  transverse dynamics of cascade approaches 
in comparison to hydrodynamical models, as pointed out by various studies 
on transverse mass spectra of different hadron species 
performed within the present framework 
\cite{Li:2008qm,Petersen:2009mz,Santini:2009nd} and by a previous 
independent comparative analysis of hadronic cascade dynamics 
vs. hydrodynamics done by Huovinen et al. \cite{Huovinen:2001cy}.
A more explicit view of the effect is shown in Fig. \ref{cascade_pt} 
where the transverse mass spectra of the three components are 
presented separately. We see that 
the contribution to dimuon emission from $\rho$ mesons that are merged
into the cascade at the transition point via the 
Cooper-Frye equation and emit in the cascade stage of the evolution
(dashed line) has flatter spectra than the one from $\rho$ mesons 
generated in the cascade stage (dotted-dashed line). 
The emission from the stage that precedes the hydrodynamical evolution presents 
as well steeper spectra than the one from $\rho$ mesons generated at the 
transition point (the latter is maximally coupled to the flow). The reason is 
that this contribution is mainly 
due to early decay of $\rho$ mesons generated by string excitation in first 
chance nucleon-nucleon collisions and therefore reminds closely the steep 
trend exhibited by dimuon transverse spectra in $pp$ collisions.

Going back to the comparison with NA60 data shown in Fig. \ref{fig1}, 
we found that the sum of the thermal and the cascade contributions 
lead to an overestimation of the peak region. 
The reason for the disagreement might be twofold:
 
\begin{itemize} 
\item As mentioned in the previous section, it can be argued that in medium modifications of the spectral 
shape of the $\rho$ meson are still important during the cascade stage 
and cannot be neglected as typically done in cascade models. 
The authors of Ref. \cite{vanHees:2007th}, e.g., 
took partially into account in-medium modifications 
of the freezeout contribution by employing, at freezeout, 
the vacuum form of the $\rho$ self-energy augmented with a width 
corresponding to
the full-width-half-maximum of the in-medium spectral
function at the freezeout conditions. 
This procedure allows to effectively account for resonance decays figuring
into the $\rho$ self energy with the net effect of depleting the yield in 
the peak region. On the contrary, if resonance decays are simply added 
channel by channel in a perturbative way, the net effect is the opposite, 
i.e. to enhance the yield in the peak region 
(this property is quite general, see e.g. discussion in 
Ref.~\cite{Knoll:1998iu,CBM:2010}). 
It might be interesting to investigate these two scenarios more in detail in 
the future.

\item It can be as well argued that the modelling of the 
freezeout time scale via hadronic cascade delivers a too slow decoupling 
of the $\rho$ meson as a consequence of a longer persistence of 
the ``pion wind'' via processes such as e.g. 
$\pi\pi\rightarrow\rho\rightarrow\pi\pi$ and 
$\pi N\rightarrow N^*/\Delta^*\rightarrow\rho N$. Concerning this point, independent 
information might be obtained from 
HBT analyses. The analysis of HBT radii of pions produced in heavy-ion 
collisions at the SPS energy regime performed within the hybrid 
approach used in this work indeed suggested that at top SPS a shorter 
duration time of the freezeout process than the one obtained from the 
inclusion of a cascade stage after the hydrodynamical evolution may be favoured 
by experimental data \cite{Alt:2007uj}. 
An HBT analysis of dilepton emission might help to shed some light on the 
decoupling time of the $\rho$ meson and presumably reveal interesting features 
on the space extension of the sources contributing to the various invariant 
mass regions. 
\end{itemize}

\section{\label{Conclusions}Summary and conclusions}
In this work we employed an integrated Boltzmann+hydrodynamics 
hybrid approach based on the Ultrarelativistic 
Quantum Molecular Dynamics  
transport model with an intermediate (3+1) dimensional hydrodynamic stage 
to analyze the hadronic contribution to the dimuon excess observed in 
In+In collisions at  $E_{\rm lab}$=158$A$ GeV. 
This is the first time that dilepton emission both in the thermal and 
non-thermal regime is 
studied within such a macro+micro hybrid approach.
We found that three regions can be identified in the dilepton invariant spectra. 
The very low mass region of the spectrum is dominated by thermal 
radiation, the region around the $\rho$ meson peak is 
dominated by late stage cascade dilepton emission and the 
intermediate region receives both contributions from hadronic and 
QGP emission, with the QGP accounting for about half of the total emission. 
The invariant mass regions $M$$<$0.5 GeV and  $M$$>$1, 
dominated by thermal radiation, show reasonable 
$p_T$ scaling revealing that if thermal rates are convoluted in a 
dynamical model a comprehensive interpretation of $M$ and $p_T$ path 
observed experimentally can be achieved. The model, however, 
fails in describing the region around the vector meson peak for low 
$p_T$, mostly 
due to the presence of a copious emission during the cascade which follows the 
hydrodynamical evolution.  The comparison to experimental data, 
seems to disfavour the presence of a long lived cascade emission
in which the $\rho$ meson can be approximated 
by its vacuum properties.  

The present calculation represents a 
first effort towards a more consistent treatment of 
the dynamics of thermal dileptons from in-medium modified 
hadrons as well as non-hadronic sources, which was so far 
entrusted to more schematic models. 
In this paper, we discussed the dimuon invariant mass region 
$2m_\mu$$<$$M$$<$1.5 GeV. The NA60 experiment, however, has succeeded in isolating the excess also in the region 1.5$<$$M$$<$2.6 GeV \cite{Arnaldi:2008er}, and a nice matching between the low and the intermediate mass region analyses was achieved \cite{Arnaldi:2008er,Damjanovic:2009zz}. 
Extension of the present approach in order to enable for the investigation of invariant masses up to 2.6 GeV is highly desirable and planned for the future. We expect it to bring further insights into the role of partonic contributions to the dimuon excess measured in the intermediate mass region.

\begin{acknowledgments}
The authors acknowledge useful 
discussions with  R.~Rapp, H.~van Hees and B.~B\"{a}uchle. 
G.~Moschelli and T.~Kolleger are thanked for their assistence in the use 
of ROOT.
We thank S.~Damjanovic for providing the experimental data.
This work was supported by the Hessen Initiative 
for Excellence (LOEWE) through the Helmholtz International Center for FAIR 
(HIC for FAIR). Computational resources have been provided by 
the Center for Scientific Computing. 
\end{acknowledgments}

\bibliography{biblio_noarxiv}

\end{document}